\documentclass[onecolumn,aps,nofootinbib]{revtex4}
\usepackage{ulem}
\usepackage{graphicx}
\usepackage{dcolumn}
\usepackage{bm}
\usepackage{subfigure}
\usepackage{slashed}
\usepackage{marvosym}
\usepackage{textcomp}
\usepackage{threeparttable}
\usepackage{booktabs}
\usepackage{setspace}
\usepackage{latexsym}
\usepackage{mciteplus}
\usepackage{amsmath,amsthm,amssymb}
\usepackage{color}

\begin{document}

\title{Magnetic field induced neutrino chiral transport near equilibrium}
\author{Naoki Yamamoto$^1$ and Di-Lun Yang$^{1,2}$}
\affiliation{$^1$Department of Physics,  Keio University, Yokohama 223-8522, Japan\\
$^2$Institute of Physics, Academia Sinica, Taipei, 11529, Taiwan}
\begin{abstract}
Based on the recently formulated chiral radiation transport theory for left-handed neutrinos, we study the chiral transport of neutrinos near thermal equilibrium in core-collapse supernovae. We first compute the near-equilibrium solution of the chiral radiation transport equation under the relaxation time approximation, where the relaxation time is directly derived from the effective field theory of the weak interaction. By using such a solution, we systematically derive analytic expressions for the nonequilibrium corrections of the neutrino energy-momentum tensor and neutrino number current induced by magnetic fields via the neutrino absorption on nucleons. In particular, we find the nonequilibrium neutrino energy current proportional to the magnetic field. We also discuss its phenomenological consequences such as the possible relation to pulsar kicks.
\end{abstract}
\maketitle
 
\section{Introduction}
One of the most important properties of neutrinos in the Standard Model of particle physics is the left-handedness. Although neutrinos are expected to play important roles in the explosion dynamics of core-collapse supernovae, this property has been neglected in the conventional neutrino radiation transport theory \cite{Lindquist1966,Castor1972,Bruenn:1985en,Mihalas,Sumiyoshi:2012za} applied so far; for recent reviews on the theoretical aspects of core-collapse supernovae, see, e.g., Refs.~\cite{Kotake:2012nd,Burrows:2012ew,Foglizzo:2015dma,Janka:2016fox,Muller:2016izw,Radice:2017kmj}. It is thus important to study the effects of chirality of neutrinos on the dynamics of the core-collapse supernova as pointed out in Ref.~\cite{Yamamoto:2015gzz}.

Recently, starting from the underlying quantum field theory, the authors of this paper have systematically constructed the neutrino radiation transport theory incorporating the effects of chirality. It is dubbed as the chiral radiation transport theory \cite{Yamamoto:2020zrs}. Unlike the conventional neutrino radiation hydrodynamics, this theory explicitly breaks the spherical symmetry and axisymmetry of the system by the quantum effects related to the chirality. Moreover, novel transport phenomena that have been missed in the conventional theory emerge, which may qualitatively change the time evolution of the system. The construction of such a theory was made possible thanks to the recent developments of the kinetic theory for chiral fermions, called chiral kinetic theory, in high-energy physics \cite{Son:2012wh,Stephanov:2012ki,Son:2012zy,Chen:2012ca,Chen:2014cla,Chen:2015gta,Hidaka:2016yjf,Hidaka:2017auj,Hidaka:2018ekt,Mueller:2017lzw,Mueller:2017arw,Huang:2018wdl,Carignano:2018gqt,Carignano:2019zsh,Lin:2019ytz,Liu:2018xip}. 

In this paper, based on this chiral radiation transport theory for neutrinos, we study the chiral transport of neutrinos near thermal equilibrium in core-collapse supernovae. 
We first compute the near-equilibrium solution of the chiral radiation transport equation under the relaxation time approximation, where the relaxation time is directly derived from the effective field theory of the weak interaction. By using this solution, we then analytically derive the nonequilibrium corrections of the neutrino energy-momentum tensor and current induced by magnetic fields through the neutrino absorption on nucleons. In particular, we find the nonequilibrium neutrino energy current and neutrino number current proportional to the magnetic field; see Eqs.~(\ref{main_T}) and (\ref{main_J}). Although the asymmetric neutrino emission induced by the strong magnetic field was also discussed in previous works in relation to the possible origin of the pulsar kicks \cite{Vilenkin:1995um,Horowitz:1997mk,Horowitz:1997fb,Roulet:1997sw,Lai:1997mm,Lai:1998sz,Arras:1998cv,Arras:1998mv,Goyal:1998nq,Kaminski:2014jda}, this work is the first, to the best of our knowledge, to derive the explicit form of the magnetic field induced energy-momentum tensor of neutrinos by systematically taking into account the effects of chirality of leptons. This work, together with our previous work \cite{Yamamoto:2020zrs}, also explicitly bridges the gap between the microscopic theory of the weak interaction for neutrinos and the neutrino radiation hydrodynamics.

The paper is organized as follows: In Sec.~\ref{sec_CRT}, we review the chiral radiation transport theory for neutrinos. In Sec.~\ref{sec_solution}, using the relaxation time derived from the effective theory of the weak interaction, we compute the near-equilibrium solution of the chiral radiation transport theory. In Sec.~\ref{sec_EM}, we derive generic expressions for the neutrino energy-momentum tensor and current near equilibrium. In Sec.~\ref{sec_nonequilbirum_B_correcions}, we compute the nonequilibrium corrections on the neutrino energy-momentum tensor and current induced by the magnetic field. Section~\ref{sec_discussion} is devoted to discussions and outlook. 

Throughout this work, we assume massless neutrinos. We use the Minkowski metric $\eta_{\mu\nu}=\text{diag}\{+,-,-,-\}$. We define the Levi-Civita tensor $\epsilon^{\mu\nu\alpha\beta}=\hat{\epsilon}^{\mu\nu\alpha\beta}/\sqrt{-g}$, where $\hat{\epsilon}^{\mu\nu\alpha\beta}$ denotes the permutation symbol and $g$ represents the determinant of the spacetime metric with the convention $\hat{\epsilon}^{0123}=-\hat{\epsilon}_{0123}=1$. For a given vector $V^{\mu}$, the unit vector is denoted by $\hat{V}^{\mu} = V^{\mu}/|{\bm V}|$ with $\bm V$ being the spatial component of $V^{\mu}$. We absorb the electric charge $e$ into the definition of the gauge field $A_{\mu}$. We also introduce the notations $A_{\{\rho}B_{\sigma\}}\equiv (A_{\rho}B_{\sigma}+A_{\sigma}B_{\rho})/2$ and  $A_{[\rho}B_{\sigma]}\equiv (A_{\rho}B_{\sigma}-A_{\sigma}B_{\rho})/2$. After Sec.~\ref{sec_CRT}, we take $\hbar = c = k_{\rm B} = 1$ except where the $\hbar$ expansion is shown.

\section{Chiral radiation transport theory for neutrinos}
\label{sec_CRT}
In this section, we review the chiral radiation transport theory for neutrinos developed in Ref.~\cite{Yamamoto:2020zrs} that will be applied in the following sections.
The general relativistic form of the chiral transfer equation with collisions for left-handed neutrinos is given by%
\footnote{In this chiral radiation transport theory, neutrinos are treated as approximately massless and all quantum effects associated with the small but finite neutrino mass are neglected.}
\begin{gather}
\label{CRT1}
\left[q^{\mu}(\partial_{\mu} - \Gamma^{\lambda}_{\mu \rho} q^{\rho} \partial_{q \lambda}) - \hbar c (D_{\mu} S^{\mu \nu}_{(n)}) \partial_{\nu} 
+ \hbar c S^{\mu \nu}_{(n)} q^{\rho} R^{\lambda}_{\rho \mu \nu} \partial_{q \lambda} \right] f^{(\nu)}_{q(n)} 
= (1-f^{(\nu)}_{q(n)})\Gamma^{<}_{q(n)}-f_{q(n)}^{(\nu)}\Gamma^{>}_{q(n)}\,, \\
\label{CRT2}
\Gamma^{\lessgtr}_{q (n)} = (q^{\nu} - \hbar c D_{\mu} S^{\mu\nu}_{q(n)}) \Sigma^{\lessgtr}_{\nu}\,.
\end{gather}
Here, $\partial_{\mu}$ and $\partial_{q\mu}$ denote the spacetime and four-momentum derivatives, respectively, $f_{q(n)}^{(\nu)} = f_{(n)}^{(\nu)}(x, q)$ is the distribution function of the left-handed neutrino which generically depends on the frame vector $n^{\mu}$ (see below), $D_{\mu} = \nabla_{\mu}-\Gamma^{\lambda}_{\mu\nu}q^{\nu}\partial_{q\lambda}$ is the horizontal lift of $\nabla_{\mu}$ defined such that $D_{\mu} q^{\nu} = 0$ with $\Gamma_{\mu \nu}^{\lambda}$ the Christoffel symbol,  $\nabla_{\mu}$ is the covariant derivative $\nabla_{\mu} V_{\nu} = \partial_{\mu} V_{\nu} - \Gamma_{\mu \nu}^{\lambda} V_{\lambda}$ for a vector $V_{\nu}$, $S^{\mu\nu}_{q(n)} = \epsilon^{\mu\nu\alpha\beta}{q_{\alpha}n_{\beta}}/({2q\cdot n})$
is the spin tensor for spin 1/2 fermions with $n^{\mu}$ the frame vector satisfying $n^2 =1$, $R^{\lambda}_{\rho\mu\nu}=2\partial_{[\mu}\Gamma^{\lambda}_{\nu]\rho}+2\Gamma^{\lambda}_{\alpha[\mu}\Gamma^{\alpha}_{\nu]\rho}$ is the Riemann tensor, and $\Sigma^{\lessgtr}_{\mu}$ are the lesser and greater self-energies. The emission and absorption rates are given by $R_{\rm emis} = \Gamma^{<}/q^0$ and $R_{\rm abs} = \Gamma^{>}/q^0$, respectively.
The terms related to the spin tensor $S^{\mu\nu}_{q(n)}$ in Eqs.~(\ref{CRT1}) and (\ref{CRT2}) that have been missed in the conventional neutrino transport theory explicitly break the spherical symmetry and axisymmetry of the system. 

Note that the dependence of the spin tensor $S^{\mu\nu}_{q(n)}$ on the frame vector $n^{\mu}$ emerges as a choice of the spin basis, and consequently, $f_{q(n)}^{(\nu)}$ and $\Gamma^{\lessgtr}_{q (n)}$ also depend on $n^{\mu}$ \cite{Chen:2015gta,Hidaka:2016yjf}. However, the physical quantities do not depend on the choice of $n^{\mu}$ at the end. Below we will always choose the frame vector $n^{\mu} = \xi^{\mu} \equiv (1,\bm 0)$ in the inertial frame, then we have $\nabla_{\mu}n_{\nu}=0$, $D_{\mu}S^{\mu\nu}_{(n)}=0$, and $R^{\lambda}_{\rho\mu\nu}=0$, and all the corrections due to the chirality of neutrinos appear in the collision term as $\Gamma^{\lessgtr}_{q} = (q^{\nu} - \hbar c S^{\mu\nu}_{q} D_{\mu}) \Sigma^{\lessgtr}_{\nu}$. Accordingly, we will not hereafter highlight the frame dependence of the quantities, such as $f^{(\nu)}_{q(n)}$. In this case, the chiral radiation transport equation reads
\begin{eqnarray}
	\label{CKT_inertial_con}
	\Box_{\rm i}f_{q}^{(\nu)}
	=\frac{1}{{E_{\rm i}}}\left[(1-f^{(\nu)}_{q})\Gamma^{<}_{q}-f_{q}^{(\nu)}\Gamma^{>}_{q} \right]\,,
\end{eqnarray}
where $\Box_{\rm i}$ is given by \cite{Sumiyoshi:2012za}
\begin{eqnarray}
\Box_{\rm i}\equiv \frac{1}{c}\partial_{t_{\rm i}}+ \frac{\mu_{\rm i}}{r^2}\partial_{r}r^2+\frac{\sqrt{1-\mu_{\rm i}^2}}{r}\left(\frac{\cos\bar{\phi}_{\rm i}}{\sin\theta}\partial_{\theta}\sin\theta+\frac{\sin\bar{\phi}_{\rm i}}{\sin\theta}\partial_{\phi}\right)
+\frac{1}{r}\partial_{\mu_{\rm i}}(1-\mu_{\rm i}^2)
-\frac{\sqrt{1-\mu_{\rm i}^2}}{r}\cot\theta\partial_{\bar{\phi}_{\rm i}}\sin\bar{\phi}_{\rm i}\,.
\end{eqnarray}
Here, we adopt the spherical coordinate system $(r, \theta, \phi)$ for the position and $(E_{\rm i}, \bar \theta_{\rm i}, \bar \phi_{\rm i})$ for the momentum of the neutrino and the subscripts ``${\rm i}$" stand for the quantities in the inertial frame. We also defined $\mu_{\rm i} \equiv \cos \bar \theta_{\rm i}$.
Note that $\Box_{\rm i}$ may also be written in a more generic form via the horizontal lift, $\Box_{\rm i}=q\cdot D/{E_{\rm i}}$.

For the collision term, we will focus on the neutrino absorption on nucleons $\nu^{\rm e}_{\rm L}(q)+{\rm n}(k)\rightleftharpoons {\rm e}_{\rm L}(q')+{\rm p}(k')$. We are interested in the length scale much larger than the mean free path in the matter sector composed of electrons and nucleons. In this case, ignoring the viscous corrections and the gradients of the temperature and chemical potentials, we may decompose $\bar \Gamma^{\lessgtr}_{q}$ as
\begin{eqnarray}
\label{ab_rad_rates}
\bar \Gamma^{\lessgtr}_{q}\approx \bar \Gamma^{(0)\lessgtr}_{q}
+\hbar \bar \Gamma^{(\omega)\lessgtr}_{q} (q\cdot \omega)
+\hbar \bar \Gamma^{(B)\lessgtr}_{q} (q\cdot B),
\label{Gamma_decomp}
\end{eqnarray}
where $\bar{O}$ stands for a quantity $O$ in local thermal equilibrium, $\omega^{\mu}=\epsilon^{\mu\nu\alpha\beta}u_{\nu} \partial_{\alpha} u_{\beta}/2$ is the vorticity, and $B^{\mu}=\epsilon^{\mu\nu\alpha\beta}u_{\nu}F_{\alpha\beta}/2$ is the magnetic field defined in the fluid rest frame with $u^{\mu}$ being the fluid four velocity and $F_{\alpha \beta}$ the field strength of the U(1) electromagnetic gauge field.
The expression for the classical term $\bar \Gamma^{(0)\lessgtr}_{q}$ was derived in Ref.~\cite{Reddy:1997yr}, while the expressions for the quantum corrections $\bar \Gamma^{(\omega)\lessgtr}_{q}$ and $\bar \Gamma^{(B)\lessgtr}_{q}$ were derived in Ref.~\cite{Yamamoto:2020zrs} based on the Fermi theory of the weak interaction under the nonrelativistic approximation for nucleons with the mass $M_{\rm n} \approx M_{\rm p} \approx M$ and the ``quasi-isoenergetic" approximation that allows for the energy transfer up to $O(1/M)$. Their explicit expressions are
\begin{eqnarray}
\nonumber
\bar \Gamma^{(0)>}_{q}&\approx& \frac{1}{\pi \hbar^4 c^4} \big(g_{\rm V}^2+3g_{\rm A}^2\big){G}_{\rm F}^2 (q\cdot u)^3(1-f^{({\rm e})}_{0,q})\left(1-\frac{3q\cdot u}{M c^2}\right)	\frac{n_{{\rm n}}-n_{\rm p}}{1-{\rm e}^{\beta(\mu_{\rm p}-\mu_{{\rm n}})}}\,,
\label{Gamma_0}
\\
\bar \Gamma^{(0)<}_{q}&\approx& \frac{1}{\pi \hbar^4 c^4} \big(g_{\rm V}^2+3g_{\rm A}^2\big){G}_{\rm F}^2 (q\cdot u)^3f^{({\rm e})}_{0,q}\left(1-\frac{3q\cdot u}{M c^2}\right)	\frac{n_{{\rm p}}-n_{\rm n}}{1-{\rm e}^{\beta(\mu_{\rm n}-\mu_{{\rm p}})}}\,,
\end{eqnarray}
\begin{eqnarray}
\nonumber
\bar \Gamma^{(B)>}_{q} &\approx& \frac{1}{2\pi \hbar^4 c^4 M} \big(g_{\rm V}^2+3g_{\rm A}^2\big){G}_{\rm F}^2 (q\cdot u)(1-f^{({\rm e})}_{0,q})\left(1-\frac{8q\cdot u}{3M c^2}\right)	\frac{n_{{\rm n}}-n_{\rm p}}{1-{\rm e}^{\beta(\mu_{\rm p}-\mu_{{\rm n}})}} \,,
\\
\bar \Gamma^{(B)<}_{q}&\approx& \frac{1}{2\pi \hbar^4 c^4 M} \big(g_{\rm V}^2+3g_{\rm A}^2\big){G}_{\rm F}^2 (q\cdot u)f^{({\rm e})}_{0,q}\left(1-\frac{8q\cdot u}{3M c^2}\right)	\frac{n_{{\rm p}}-n_{\rm n}}{1-{\rm e}^{\beta(\mu_{\rm n}-\mu_{{\rm p}})}} \,,
\label{Gamma_B}
\end{eqnarray}
\begin{eqnarray}
\nonumber
\bar \Gamma^{(\omega)>}_{q}&\approx & \frac{1}{2\pi \hbar^4 c^4} \big(g_{\rm V}^2+3g_{\rm A}^2\big){G}_{\rm F}^2 (q\cdot u)^2(1-f^{({\rm e})}_{0,q})\left(\frac{2}{E_{\rm i}}+\beta  f^{({\rm e})}_{0,q}\right)
\frac{n_{{\rm n}}-n_{\rm p}}{1-{\rm e}^{\beta(\mu_{\rm p}-\mu_{{\rm n}})}}\,,
\\
\bar \Gamma^{(\omega)<}_{q}&\approx & \frac{1}{2\pi \hbar^4 c^4} \big(g_{\rm V}^2+3g_{\rm A}^2\big){G}_{\rm F}^2 (q\cdot u)^2f^{({\rm e})}_{0,q}\left(\frac{2}{E_{\rm  i}}-\beta (1-f^{({\rm e})}_{0,q})\right)
\frac{n_{{\rm p}}-n_{\rm n}}{1-{\rm e}^{\beta(\mu_{\rm n}-\mu_{{\rm p}})}}\,,
\label{Gamma_omega}
\end{eqnarray}
where $G_{\rm F}$ is the Fermi constant and $g_{\rm V}=1$ and $g_{\rm A} \approx 1.27$ are the nucleon vector and axial charges, respectively. We also introduced the Fermi-Dirac distributions
\begin{eqnarray}
f^{(i)}_{0,q}=\frac{1}{{\rm e}^{\beta (q\cdot u-\mu_i)}+1} \qquad (i={\rm n}, {\rm p}, {\rm e})\,,
\end{eqnarray}
where $\beta=1/(k_{\rm B}T)$ with $T$ being temperature and $\mu_i$ chemical potentials for $i={\rm n}, {\rm p}, {\rm e}$, and $n_{{\rm n}/{\rm p}} = \int \frac{{\rm d}^3 {\bm k}}{(2\pi \hbar)^{3}} f^{({\rm n}/{\rm p})}_{0,k}$ are neutron/proton densities.

Although $q\cdot u\approx E_{\rm i}\equiv q\cdot \xi$ for the on-shell fermions, we rigorously distinguish between $q\cdot u$ and ${E_{\rm i}}$ in the expressions of $\bar \Gamma^{\lessgtr}_{q}$ above. This difference will become important in computing the neutrino energy-momentum tensor $T^{\mu \nu}_{(\nu)}$ and neutrino current $J^{\mu}_{(\nu)}$ below since $\nabla_{\mu}(q\cdot u)\neq \nabla_{\mu}{E_{\rm i}}=0$.

For a given $f^{(\nu)}_{q}$, the energy-momentum tensor and current of neutrinos are given by \cite{Yamamoto:2020zrs}
\begin{align}
\label{T_nu}
T^{\mu\nu}_{(\nu)}&=\int_q 4\pi\delta(q^2)\Big(q^{\mu}q^{\nu}f^{(\nu)}_{q}-\hbar c q^{\{\mu}S_q^{\nu\}\rho}\mathcal{D}_{\rho}f^{(\nu)}_{q}\Big)\,,
\\
\label{J_nu}
J^{\mu}_{(\nu)}& =\int_q 4\pi\delta(q^2)\Big(q^{\mu} f^{(\nu)}_{q}-\hbar c S_q^{\mu \rho}\mathcal{D}_{\rho}f^{(\nu)}_{q}\Big)\,,
\end{align}
where $\mathcal{D}_{\mu}f^{(\nu)}_{q} \equiv D_{\mu}f^{(\nu)}_{q}-\mathcal{C}_{\mu}[f^{(\nu)}_{q}]$ with $\mathcal{C}_{\mu}[f^{(\nu)}_{q}]\equiv\Sigma_{\mu}^{<}(1-f^{(\nu)}_{q})-\Sigma_{\mu}^{>}f^{(\nu)}_{q}$ and we introduced the notation (with setting $\sqrt{-g}=1$ in flat spacetime)
\begin{eqnarray}
\int_q \equiv \frac{1}{\hbar^3}\int\frac{{\rm d}^4q}{(2\pi)^4}\,.
\end{eqnarray}
The energy-momentum transfer from neutrino radiation to matter is dictated by the energy-momentum conservation law
\begin{eqnarray}
\label{cons_EM0}
	\nabla_{\mu}T^{\mu\nu}_{\text{mat}}=-\nabla_{\mu}T^{\mu\nu}_{(\nu)}\,,
\end{eqnarray}
where $T^{\mu\nu}_{\text{mat}}$ is the energy-momentum tensor of the matter sector composed of electrons, neutrons, and protons. In the presence of the electromagnetic fields, the energy-momentum conservation law is modified to
\begin{eqnarray}
\label{cons_EM}
\nabla_{\mu}T^{\mu\nu}_{\text{mat}}=F^{\nu\mu}\big(J_{({\rm p})\mu}-J_{({\rm e})\mu}\big)-\nabla_{\mu}T^{\mu\nu}_{(\nu)}\,,
\end{eqnarray}
where $J_{(\rm p)\mu}$ is the electric current of protons and $J^{\mu}_{(\rm e)}=J^{\mu}_{\rm R(\rm e)}+J^{\mu}_{\rm L(\rm e)}$ is the electric current of electrons including the contributions from both right- and left-handed electrons. 

In addition, we also have the lepton current conservation, anomaly relation for the axial current, electric current conservation, and baryon current conservation, which are given by
\begin{gather}
\label{cons_lept}
\nabla_{\mu}J^{\mu}_{(\rm e)}+\nabla_{\mu}J^{\mu}_{(\nu)}=0\,,
\\
\label{cons_axial}
\nabla_{\mu}J^{\mu}_{5(\rm e)}-\nabla_{\mu}J^{\mu}_{(\nu)}=-\frac{1}{2\pi^2 \hbar^2} E\cdot B\,,
\\
\label{cons_echarge}
\nabla_{\mu}J^{\mu}_{({\rm p})}-\nabla_{\mu}J^{\mu}_{({\rm e})}=0\,,
\\
\label{cons_baryon}
\nabla_{\mu}J^{\mu}_{({\rm p})}+\nabla_{\mu}J^{\mu}_{({\rm n})}=0\,,
\end{gather}
respectively, where $J^{\mu}_{5(\rm e)}=J^{\mu}_{\rm R(\rm e)}-J^{\mu}_{\rm L(\rm e)}$ is the axial current of electrons, $J^{\mu}_{({\rm n})}$ is the current of neutrons, and $E^{\mu}=F^{\mu\nu}u_{\nu}$ is the electric field defined in the fluid rest frame. When the matter sector is in equilibrium, its state is characterized by $u^{\mu}$, $T$, $\mu_{\rm p}$, $\mu_{\rm n}$, the electron (vector) chemical potential $\mu_{\rm e}=(\mu_{\rm eR} + \mu_{\rm eL})/2$ and electron chiral chemical potential $\mu_{{\rm e}5}=(\mu_{\rm eR} - \mu_{\rm eL})/2$.%
\footnote{When the finite electron mass $m_{\rm e}$ is taken into account, it attenuates $\mu_{{\rm e}5}$ by the chirality flipping process \cite{Grabowska:2014efa}. However, the following discussion and our main results will not be affected even when $\mu_{{\rm e}5} = 0$, and the effects of the electron mass on our results can be treated as a perturbation in terms of $m_{\rm e}/\mu \ll 1$.}

So far, the governing equations are generic and are applicable even when the neutrino sector is far away from equilibrium. In the following, we will consider the case where the neutrino sector is near equilibrium (which is the case near the core of the supernova), and then its evolution is further characterized by the neutrino chemical potential $\mu_{\nu}$. Here, for simplicity, we assume that the matter sector and neutrino sector have the same temperature and fluid velocity.
In this case, the time evolution of the system, when ignoring the evolution of the dynamical electromagnetic fields, is governed by Eqs.~(\ref{cons_EM0}) and (\ref{cons_lept})--(\ref{cons_baryon}). In total, one has nine variables and eight conservative equations. To form a closure for the equations and variables, we have to incorporate the $\beta$ equilibrium condition, $\mu_{\rm e}+\mu_{\rm p}=\mu_{\nu}+\mu_{\rm n}$. In the presence of dynamical electromagnetic fields, we need to solve Eqs.~(\ref{cons_EM}) and (\ref{cons_lept})--(\ref{cons_baryon}) coupled to Maxwell's equation simultaneously.

\section{Near-equilibrium solution for the chiral transport equation}
\label{sec_solution}
Based on the chiral radiation transport equation above, let us solve for the near-equilibrium distribution function of neutrinos.
In the following, we take $\hbar = c = k_{\rm B} = 1$ except where the $\hbar$ expansion is shown.

We first consider the case of equilibrium state for neutrinos where the collision term vanishes,
\begin{eqnarray}
\label{KMS_cond}
(1-\bar{f}^{(\nu)}_{q})\bar{\Gamma}^{<}_{q}=\bar{f}^{(\nu)}_{q}\bar{\Gamma}^{>}_{q}.
\end{eqnarray}
We decompose the neutrino distribution function as $\bar{f}^{(\nu)}_{q}=f^{(\nu)}_{0,q}+\hbar f^{(\nu)}_{1,q}$, where $\hbar f^{(\nu)}_{1,q}$ denotes the quantum correction on the classical distribution function in equilibrium, $f^{(\nu)}_{0,q}$. It then follows that
\begin{eqnarray}
\frac{\bar{\Gamma}^{>}_{q} }{\bar{\Gamma}^{<}_{q} } \approx \frac{1-f^{(\nu)}_{0,q}}{f^{(\nu)}_{0,q}}\left[1-\frac{\hbar f^{(\nu)}_{1,q}}{f^{(\nu)}_{0,q}(1-f^{(\nu)}_{0,q})}\right]\,.
\end{eqnarray}
From Eqs.~(\ref{Gamma_decomp})--(\ref{Gamma_omega}) on the other hand, we have
\begin{eqnarray}\nonumber
\label{Gamma_ratio}
\frac{\bar{\Gamma}^{>}_{q} }{\bar{\Gamma}^{<}_{q} }&\approx&\frac{\bar \Gamma^{(0)>}_{q}}{\bar \Gamma^{(0)<}_{q}}\left[1+\hbar(q\cdot \omega)\left(\frac{\bar{\Gamma}^{(\omega)>}_{q}}{\bar \Gamma^{(0)>}_{q}}-\frac{\bar{\Gamma}^{(\omega)<}_{q}}{\bar \Gamma^{(0)<}_{q}}\right)
+\hbar(q\cdot B)\left(\frac{\bar{\Gamma}^{(B)>}_{q}}{\bar \Gamma^{(0)>}_{q}}-\frac{\bar{\Gamma}^{(B)<}_{q}}{\bar \Gamma^{(0)<}_{q}}\right)
\right]
\\
&=&-\frac{(1-f^{({\rm e})}_{0,q})\big(1-{\rm e}^{\beta(\mu_{\rm n}-\mu_{{\rm p}})}\big)}{f^{({\rm e})}_{0,q}\big(1-{\rm e}^{\beta(\mu_{\rm p}-\mu_{{\rm n}})}\big)}
\left[1+\hbar\frac{\beta q\cdot \omega}{2{q\cdot u}}\left(1-\frac{3{q\cdot u}}{M}\right)^{-1}
\right]
\end{eqnarray}
up to $O(\hbar)$.
Comparing the right-hand sides of Eqs.~(\ref{KMS_cond}) and (\ref{Gamma_ratio}) order by order in $\hbar$, we obtain
\begin{gather}
\label{f_0}
\frac{1-f^{(\nu)}_{0,q}}{f^{(\nu)}_{0,q}}=-\frac{(1-f^{({\rm e})}_{0,q})\big(1-{\rm e}^{\beta(\mu_{\rm n}-\mu_{{\rm p}})}\big)}{f^{({\rm e})}_{0,q}\big(1-{\rm e}^{\beta(\mu_{\rm p}-\mu_{{\rm n}})}\big)}\,, \\
\label{f_1}
f^{(\nu)}_{1,q}=-f^{(\nu)}_{0,q}(1-f^{(\nu)}_{0,q})\frac{\beta q\cdot \omega}{2{q\cdot u}}\left(1-\frac{3{q\cdot u}}{M}\right)^{-1}
=\frac{\beta q\cdot \omega}{2{q\cdot u}}\left(1-\frac{3{q\cdot u}}{M}\right)^{-1}\partial_{\beta q\cdot u}f^{(\nu)}_{0,q}\,.
\end{gather}
We accordingly obtain the equilibrium distribution function for neutrinos,
\begin{eqnarray}
\bar{f}^{(\nu)}_{q}=\frac{1}{{\rm e}^{h}+1}\,,
\end{eqnarray}
where
\begin{eqnarray}
h \approx \beta(q\cdot u-\mu_{\nu})+\hbar\beta\frac{q\cdot\omega}{2q\cdot u}+O\left(\frac{q\cdot u}{M}\right)\,,
\end{eqnarray}
with $\mu_{\nu}$ the neutrino chemical potential that satisfies the $\beta$ equilibrium condition $\mu_{\rm e}+\mu_{\rm p}=\mu_{\nu}+\mu_{\rm n}$.
For consistency, we here drop the $q\cdot u/M$ correction since the $O(1/M)$ corrections on $\bar \Gamma^{(\omega)\lessgtr}_{q}$ are already neglected based on the nonrelativistic approximation above. After dropping this term, $\bar{f}^{(\nu)}_{q}$ above agrees with the equilibrium distribution function in Refs.~\cite{Chen:2015gta,Hidaka:2017auj}. 

When neutrinos are not in complete equilibrium but are close to equilibrium, we may rewrite the collision term in the relaxation time approximation,
\begin{eqnarray}
\frac{1}{{E_{\rm i}}}\left[(1-f^{(\nu)}_{q})\Gamma^{<}_{q}-f_{q}^{(\nu)}\Gamma^{>}_{q} \right]\approx -\frac{\delta f^{(\nu)}_{q}}{\tau}\,,
\end{eqnarray}  
where $\delta f^{(\nu)}_{q}\equiv f^{(\nu)}_{q}-\bar{f}^{(\nu)}_{q}$ is the fluctuation of the distribution function and  $\tau={E_{\rm i}}/\big(\bar{\Gamma}^{>}_{q}+\bar{\Gamma}^{<}_{q}\big)$ denotes a momentum-dependent relaxation time which describes how long the system returns to the equilibrium state. From Eqs.~(\ref{KMS_cond}) and (\ref{ab_rad_rates}), we find
\begin{eqnarray}\label{relax_time}
\tau\approx \frac{E_{\rm i}(1-\bar{f}_{q}^{(\nu)})}{\bar{\Gamma}^{(0)>}_{q}}\left[1
-\hbar \frac{\bar \Gamma^{(\omega)>}_{q} (q\cdot \omega)}{\bar \Gamma^{(0)>}_{q}}
-\hbar \frac{\bar \Gamma^{(B)>}_{q} (q\cdot B)}{\bar \Gamma^{(0)>}_{q}}\right]\,.
\end{eqnarray}
Solving Eq.~(\ref{CKT_inertial_con}), the perturbative solution of $\delta f^{(\nu)}_{q}$ is given by 
\begin{eqnarray}\label{delf_sol}
\delta f^{(\nu)}_{q}\approx -\tau\Box_{\rm i}\bar{f}_{q}^{(\nu)}=-\tau\frac{q\cdot D}{{E_{\rm i}}}\bar{f}_{q}^{(\nu)}\,.
\end{eqnarray}

By decomposing the relaxation time as $\tau=\tau^{(0)}+\hbar \tau^{(1)}$ via the $\hbar$ expansion, we have more explicit expressions
\begin{align}
\tau^{(0)}&= \frac{{E_{\rm i}}(1-f^{({\rm \nu})}_{0,q})}{\bar{\Gamma}^{(0)>}_{q}}=
\frac{\kappa E_{\rm i} (1-f_{0,q}^{(\nu)})}{(q\cdot u)^3 (1-f^{({\rm e})}_{0,q}) }\,, \\
\tau^{(1)}&= -\frac{{E_{\rm i}}}{\bar{\Gamma}^{(0)>}_{q}}\left[f^{({\rm \nu})}_{1,q}+(1-f^{({\rm \nu})}_{0,q})\frac{\bar{\Gamma}^{(\omega)>}_{q} (q\cdot \omega)+\bar{\Gamma}^{(B)>}_{q} (q\cdot B)}{\bar \Gamma^{(0)>}_{q}}\right] \nonumber \\
&=-\tau^{(0)}\left[\left(\frac{2}{E_{\rm  i}}+\beta \big(f^{({\rm e})}_{0,q}-f^{({\nu})}_{0,q}\big)\right)\frac{q\cdot \omega}{2q\cdot u}+\frac{q\cdot B}{2M(q\cdot u)^2}\right]\,,
\end{align}
with
\begin{align}\label{def_kappa}
\kappa \equiv \frac{\pi}{G_{\rm F}^2 (g_{\rm V}^2 + 3 g_{\rm A}^2) \delta n}\,, \qquad
\delta n \equiv \frac{n_{{\rm p}}-n_{\rm n}}{1-{\rm e}^{\beta(\mu_{\rm n}-\mu_{{\rm p}})}}\,.
\end{align}
Here, we used Eq.~(\ref{f_1}) with dropping the $O(1/M)$ terms. Note that the relaxation time is directly derived from the Fermi theory, which is the low-energy effective field theory of the weak interaction.

Some remarks are in order here.
First, one may attempt to include the magnetic moments of nucleons neglected in Ref.~\cite{Yamamoto:2020zrs}. Naively, we may take into account the effects of the nucleon magnetic moment by consistently replacing $\mu_{i}$ by $\mu_{i}-s_{i}\lambda_{i}|\bm B|/(2M)$ for $i={\rm n,p}$, where $\lambda_{i}/(2M)$ is the magnetic moment and $s_{i}=\pm 1$ denotes the spin up or down. This amounts to the replacement of $f^{(i)}_{0,k}$ by
\begin{eqnarray}
f^{(i)}_{k}\approx \frac{1}{{\rm e}^{\beta [M-\mu_{i}+\hbar\lambda_{i}|\bm B|/(2M)]}+1}
\approx f^{(i)}_{0,k}\left(1-\hbar \frac{s_{i}\lambda_{i}|\bm B|}{2MT}\right)\,,
\end{eqnarray}
for $M-\mu_{i}\gg T$ and $|\bm B|\ll MT$. In such a case, one obtains an extra contribution from the magnetic field to the relaxation time, $\tau=\tau^{(0)}+\hbar \tau^{(1)}+\hbar \delta\tau^{(0)}$, where
\begin{eqnarray}
\delta \tau^{(0)}=\tau^{(0)}\frac{|\bm B|}{2MT}\sum_{s_{\rm p},s_{\rm n}}\bigg(\frac{s_{\rm p}\lambda_{\rm p}n_{\rm p}-s_{\rm n}\lambda_{\rm n}n_{\rm n}}{n_{\rm p}-n_{\rm n}}+\frac{s_{\rm n}\lambda_{\rm n}-s_{\rm p}\lambda_{\rm p}}{{\rm e}^{\beta(\mu_{\rm p}-\mu_{\rm n})}-1}\bigg)\,.
\end{eqnarray}
In this approximation, however, the nucleon wave functions do not include the magnetic field corrections. Hence, a more systematic inclusion of the magnetic field corrections in the nucleon Wigner functions (in addition to the distribution functions) would be necessary.%
\footnote{In previous works, e.g., in Ref.~\cite{Arras:1998mv}, the nucleon magnetic moment is included in nucleon response functions. However, the scattering matrix element of polarized nucleons in vacuum is simply used, and the effects of the medium and the magnetic field on the scattering matrix element are not fully taken into account.} 
For this reason, we do not consider the magnetic moment contributions from nucleons in the present paper.

Second, one may also consider the elastic neutrino-nucleon scattering $\nu^{\ell}_{\rm L}(q)+{\rm N}(k)\rightleftharpoons \nu^{\ell}_{\rm L}(q')+{\rm N}(k')$. Nevertheless, an analytic form for the collision term in the relaxation time approximation linear to $\delta f^{(\nu)}_{q}$ cannot be derived by simply adopting the isoenergetic approximation. In light of Ref.~\cite{Yamamoto:2020zrs}, the collision term reads
\begin{eqnarray}
(1-f^{(\nu)}_{q})\Gamma^{(\text{el})<}_{q}-f^{(\nu)}_{q}\Gamma^{(\text{el})>}_{q}
=\int_p \delta(q'^2)q^{\mu} {\Pi}^{({\rm NN})}_{p,\mu\lambda}q'^{\lambda}\Big(f^{(\nu)}_{q'}-f^{(\nu)}_{q}\Big)\bigg|_{q'=q-p}+O(\hbar)\,,
\end{eqnarray}  
where the $O(\hbar)$ terms are dropped here. (The detailed structure of ${\Pi}^{({\rm NN})}_{p, \mu \lambda}$ obtained from the isoenergetic approximation can be found there.) When neutrinos are near equilibrium, one finds $f^{(\nu)}_{q-p}-f^{(\nu)}_{q}\approx \delta f^{(\nu)}_{q-p}-\delta f^{(\nu)}_{q}$ given $p\cdot u\approx |\bm p|^2/(2M)\ll q\cdot u$, where $\bm p$ is the momentum transfer. To obtain a nonvanishing collision term analytically, a further assumption for the hierarchy between the neutrino momentum $|{\bm q}|$ and the momentum transfer $|{\bm p}|$ has to be imposed. Moreover, it is necessary to consistently incorporate $O(|\bm p|/M)$ corrections and the recoil momenta on nucleons, which are already neglected in the isoenergetic approximation. Therefore, we also do not include the elastic neutrino-nucleon scattering in the present work for consistency.

\section{Neutrino energy-momentum tensor and current}
\label{sec_EM}
Given the near-equilibrium solution for $f^{(\nu)}_{q}$, we are now able to evaluate $T^{\mu \nu}_{(\nu)}$ and $J^{\mu}_{(\nu)}$ according to Eqs.~(\ref{T_nu}) and (\ref{J_nu}). 
For neutrinos near local thermal equilibrium, we decompose 
\begin{eqnarray}
T^{\mu\nu}_{(\nu)}=\bar T^{\mu\nu}_{(\nu)}+\delta T^{\mu\nu}_{(\nu)}\,,
\qquad
J^{\mu}_{(\nu)} = \bar J^{\mu}_{(\nu)}+\delta J^{\mu}_{(\nu)},
\end{eqnarray}
where 
\begin{align}
\bar T^{\mu\nu}_{(\nu)} &\equiv \int_q 4\pi\delta(q^2)\Big(q^{\mu}q^{\nu}\bar{f}^{(\nu)}_{q}-\hbar q^{\{\mu}S_q^{\nu\}\rho}D_{\rho}\bar{f}^{(\nu)}_{q}\Big)\,, \\
\label{delTrad}
\delta T^{\mu\nu}_{(\nu)}&\equiv \int_q 4\pi\delta(q^2)\Big(q^{\mu}q^{\nu}\delta f^{(\nu)}_{q}-\hbar q^{\{\mu}S_q^{\nu\}\rho}D_{\rho}\delta f^{(\nu)}_{q}\Big)\,,
\end{align}
and 
\begin{align}
\bar J^{\mu}_{(\nu)}&\equiv \int_q 4\pi\delta(q^2)\Big(q^{\mu}\bar{f}^{(\nu)}_{q}-\hbar S_q^{\mu\rho}D_{\rho}\bar{f}^{(\nu)}_{q}\Big)\,, \\
\label{delJrad}
\delta J^{\mu}_{(\nu)}&\equiv \int_q 4\pi\delta(q^2)\Big(q^{\mu}\delta f^{(\nu)}_{q}-\hbar S_q^{\mu\rho}D_{\rho}\delta f^{(\nu)}_{q}\Big)\,.
\end{align}
Note that $\mathcal{C}_{\rho}[f^{(\nu)}_{q}]\propto u_{\rho}$ up to $O(\hbar^0)$ with the matter sector in equilibrium, for which we have $\hbar S^{\nu\rho}_q\mathcal{C}_{\rho}[f^{(\nu)}_{q}] = 0$. Hence, $\hbar S^{\nu\rho}_q\mathcal{C}_{\rho}[f^{(\nu)}_{q}] = O(\hbar^2)$ and it is neglected above. 

Given $\bar{f}^{(\nu)}_{q}$, we may rewrite $\bar T^{\mu\nu}_{(\nu)}$ and $\bar J^{\mu}_{(\nu)}$ as \cite{Hidaka:2016yjf}
\begin{align}
\bar T^{\mu\nu}_{(\nu)}&=\int_q 4\pi\delta(q^2)q^{\{\mu}\Big[q^{\nu\}}-\frac{\hbar}{2}\beta \big(\omega^{\nu\}}q\cdot u-u^{\nu\}}q\cdot \omega\big)(1-f^{(\nu)}_{0,q})\Big]f^{(\nu)}_{0,q}\,,
\\
\bar J^{\mu}_{(\nu)}&=\int_q 4\pi\delta(q^2)\Big[q^{\mu}-\frac{\hbar}{2}\beta \big(\omega^{\mu}q\cdot u-u^{\mu}q\cdot \omega\big)(1-f^{(\nu)}_{0,q})\Big]f^{(\nu)}_{0,q}\,,
\end{align}
which lead to
\begin{align}
\bar T^{\mu\nu}_{(\nu)}&=\epsilon_{(\nu)}u^{\mu}u^{\nu}-p_{(\nu)}\Delta^{\mu\nu}+\hbar \xi_{\omega(\nu)}\big(\omega^{\mu}u^{\nu}+\omega^{\nu}u^{\mu}\big)\,,
\\
\bar J^{\mu}_{(\nu)}&=N_{(\nu)}u^{\mu}+\hbar \sigma_{\omega(\nu)}\omega^{\mu}\,,
\end{align}
where $\Delta^{\mu \nu} \equiv \eta^{\mu \nu} - u^{\mu} u^{\nu}$.
When $\mu_{\nu}\gg T$, we find
\begin{gather}
\epsilon_{(\nu)}=3p_{(\nu)} \approx \frac{\mu_{\nu}^4}{8\pi^2}+\frac{\mu_{\nu}^2 T^2}{4}+\frac{7\pi^2}{120}T^4\,, 
\quad
\xi_{\omega(\nu)} \approx -\bigg(\frac{\mu_{\nu}^3}{6\pi^2}+\frac{\mu_{\nu}T^2}{6}\bigg)\,.
\\
N_{(\nu)} \approx \frac{\mu_{\nu}^3}{6\pi^2} + \frac{\mu_{\nu} T^2}{6}\,,
\quad
\sigma_{\omega(\nu)} \approx -\left(\frac{\mu_{\nu}^2}{4\pi^2}+\frac{T^2}{12}\right)\,.
\label{CVE}
\end{gather}
In this case, the contribution of antineutrinos is suppressed and the transport coefficients $\xi_{\omega(\nu)}$ and $\sigma_{\omega(\nu)}$ agree with those in the chiral fluid including the contributions of both fermions and antifermions \cite{Son:2009tf,Landsteiner:2011iq,Gao:2012ix,Chen:2015gta,Yang:2018lew}. 
In particular, $\bar J^{\mu}_{(\nu)} \propto \omega^{\mu}$ above is known as the chiral vortical effect \cite{Vilenkin:1979ui,Erdmenger:2008rm,Banerjee:2008th,Son:2009tf,Landsteiner:2011cp} and Eq.~(\ref{CVE}) correctly reproduces its transport coefficient.
Although the isoenergetic approximation breaks down at $\mu_{\nu}\gg T$ \cite{Reddy:1997yr}, we assume sufficiently large $\mu_{\nu}$ such that the antineutrino distribution function is comparatively negligible yet the isoenergetic approximation is still valid.

On the other hand, inserting Eq.~(\ref{delf_sol}) into Eqs.~(\ref{delTrad}) and (\ref{delJrad}), the nonequilibrium corrections for the neutrino energy-momentum tensor and current become
\begin{align}
\label{delT_in_delf}
\delta T^{\mu\nu}_{(\nu)}&= -\int_q \frac{4\pi\delta(q^2)}{{E_{\rm i}}}\Big[q^{\mu}q^{\nu}(\tau^{(0)}+\hbar \tau^{(1)})-\hbar q^{\{ \mu}S_q^{\nu \} \rho}D_{\rho}\tau^{(0)}\Big] q\cdot D\bar{f}_{q}^{(\nu)}\,,
\\
\label{delJ_in_delf}
\delta J^{\mu}_{(\nu)}&= -\int_q \frac{4\pi\delta(q^2)}{{E_{\rm i}}}\Big[q^{\mu}(\tau^{(0)}+\hbar \tau^{(1)})-\hbar S_q^{\mu\rho}D_{\rho}\tau^{(0)}\Big] q\cdot D\bar{f}_{q}^{(\nu)}\,.
\end{align}
As $D_{\mu}$ is defined such that $D_{\mu} q^{\nu} = 0$, it follows that $D_{\mu}\mathcal{F}(q\cdot u)=\nabla_{\mu}\mathcal{F}(q\cdot u)$ for an arbitrary function $\mathcal{F}(q\cdot u)$. Accordingly, we may replace $D_{\mu}$ by $\nabla_{\mu}$ when it acts on $\tau^{(0)}$ or $f^{(\nu)}_{0,q}$ in Eqs.~(\ref{delT_in_delf}) and (\ref{delJ_in_delf}). 

We now make a further decomposition, $\delta T^{\mu\nu}_{(\nu)}=\delta T^{(0)\mu\nu}_{(\nu)}+\hbar \delta T^{(1)\mu\nu}_{(\nu)}$, where
\begin{align}
\delta T^{(0)\mu\nu}_{(\nu)}&=-\int_q \frac{4\pi\delta(q^2)}{{E_{\rm i}}}q^{\mu}q^{\nu}\tau^{(0)} q\cdot \nabla f^{(\nu)}_{0,q}\,,
\\
\delta T^{(1)\mu\nu}_{(\nu)}&= -\int_q \frac{4\pi\delta(q^2)}{{E_{\rm i}}}\Big[\tau^{(0)}q^{\{\mu}\Big(q^{\nu\}} q\cdot D f^{(\nu)}_{1,q}-S_q^{\nu\}\rho}D_{\rho}\big(q\cdot \nabla  f^{(\nu)}_{0,q}\big)\Big)+q^{\{\mu}\Big(q^{\nu\}}\tau^{(1)}- S_q^{\nu\}\rho}\big(\nabla_{\rho}\tau^{(0)}\big)\Big) q\cdot \nabla f^{(\nu)}_{0,q}
\Big]
\end{align}
correspond to the explicit classical and quantum fluctuations, respectively. However, as will be discussed later, from the $\hbar$ corrections encoded in hydrodynamic equations of motion, $\delta T^{(0)\mu\nu}_{(\nu)}$ can also yield quantum corrections comparable to $\delta T^{(1)\mu\nu}_{(\nu)}$. Similarly, we decompose as $\delta J^{\mu}_{(\nu)}=\delta J^{(0)\mu}_{(\nu)}+\hbar \delta J^{(1)\mu}_{(\nu)}$, where
\begin{align}
\delta J^{(0)\mu}_{(\nu)}&=-\int_q \frac{4\pi\delta(q^2)}{{E_{\rm i}}}q^{\mu}\tau^{(0)} q\cdot \nabla f^{(\nu)}_{0,q}\,,
\\
\delta J^{(1)\mu}_{(\nu)}&= -\int_q \frac{4\pi\delta(q^2)}{{E_{\rm i}}}\Big[\tau^{(0)} \Big(q^{\mu} q\cdot D f^{(\nu)}_{1,q}-S_q^{\mu \rho}D_{\rho}\big(q\cdot \nabla  f^{(\nu)}_{0,q}\big)\Big)+ \Big(q^{\mu}\tau^{(1)}- S_q^{\mu \rho}\big(\nabla_{\rho}\tau^{(0)}\big)\Big) q\cdot \nabla f^{(\nu)}_{0,q}
\Big]\,.
\end{align}

\section{Nonequilibrium corrections from magnetic fields}
\label{sec_nonequilbirum_B_correcions}
In this section, we derive the explicit forms of nonequilibrium corrections on the neutrino energy-momentum tensor and current. Using
\begin{eqnarray}
\nonumber
	\nabla_{\rho}f^{(\nu)}_{0,q}&=&-f^{(\nu)}_{0,q}\big(1-f^{(\nu)}_{0,q}\big)\big(q^{\lambda}\nabla_{\rho}(\beta u_{\lambda})-\nabla_{\rho}\bar{\mu}_{\nu}\big)\,,
	\\\nonumber
	D_{\rho}f^{(\nu)}_{1,q}&=&-\frac{f^{(\nu)}_{0,q}\big(1-f^{(\nu)}_{0,q}\big)}{2q\cdot u}\left[\Big(q^{\lambda}\nabla_{\rho}(\beta \omega_{\lambda})-\frac{\beta q\cdot\omega}{q\cdot u}q^{\lambda}\nabla_{\rho}u_{\lambda}\Big)
	-(1-2f^{(\nu)}_{0,q})\beta q\cdot\omega\big(q^{\lambda}\nabla_{\rho}(\beta u_{\lambda})-\nabla_{\rho}\bar{\mu}_{\nu}\big)\right]\,,
	\\
	\nabla_{\rho}\tau^{(0)}&=&\tau^{(0)}\left[-\frac{3q^{\lambda}\nabla_{\rho}u_{\lambda}}{q\cdot u}+f_{0,q}^{(\nu)}\big(q^{\lambda}\nabla_{\rho}(\beta u_{\lambda})-\nabla_{\rho}\bar{\mu}_{\nu}\big)-f_{0,q}^{(\rm e)}\big(q^{\lambda}\nabla_{\rho}(\beta u_{\lambda})-\nabla_{\rho}\bar{\mu}_{\rm e}\big)
	-\nabla_{\rho}\ln\delta n\right]\,,
\end{eqnarray}
where $\bar{\mu}_{i}\equiv \beta\mu_{i}$ for $i=\nu,{\rm e,p,n}$, we can evaluate $\delta T^{\mu\nu}_{(\nu)}$ and $\delta J^{\mu}_{(\nu)}$ explicitly. 
[Note again that the difference between $q\cdot u$ and ${E_{\rm i}}$ is essential here since $\nabla_{\mu}(q\cdot u)\neq \nabla_{\mu}{E_{\rm i}}=0$.]
Nonetheless, the full $\delta T^{\mu\nu}_{(\nu)}$ and $\delta J^{\mu}_{(\nu)}$ are rather complicated, and here we will focus on the contributions due to magnetic fields in which $\tau^{(1)}$ is involved.

In principle, the leading-order corrections $\delta T^{(0)\mu\nu}_{(\nu)}$ and $\delta J^{(0)\mu}_{(\nu)}$ may also incorporate magnetic field corrections through the hydrodynamic equations of motion that determine the temporal derivatives on thermodynamic parameters up to $O(\hbar)$. Nevertheless, as will be shown in Sec.~\ref{sec_hydro_EOM}, the possible contributions are proportional to $B\cdot\nabla_{\perp}T$ and $B\cdot\nabla_{\perp}\mu$, which are different from the forms of the viscous corrections originating from $\tau^{(1)}$ that we are interested in here. For the magnetic field induced corrections involving $\tau^{(1)}$, we find
\begin{align}
\delta T^{(1)\mu\nu}_{(\nu)B}&=-
 \int_q \frac{4\pi\delta(q^2)}{{E_{\rm i}}}q^{\mu}q^{\nu}\tau^{(0)} \frac{q\cdot B}{2M(q\cdot u)^2}f^{(\nu)}_{0,q}\big(1-f^{(\nu)}_{0,q}\big)\big(q^{\rho}q^{\lambda}{\Theta}_{\rho\lambda}-q\cdot\nabla \bar{\mu}_{\nu}\big)\,, \\
\delta J^{(1)\mu}_{(\nu)B}&=-
\int_q \frac{4\pi\delta(q^2)}{{E_{\rm i}}}q^{\mu}\tau^{(0)} \frac{q\cdot B}{2M(q\cdot u)^2}f^{(\nu)}_{0,q}\big(1-f^{(\nu)}_{0,q}\big)\big(q^{\rho}q^{\lambda}{\Theta}_{\rho\lambda}-q\cdot\nabla \bar{\mu}_{\nu}\big)\,,
\end{align}
where ${\Theta}_{\rho\lambda} \equiv \nabla_{\{\rho} \beta u_{\lambda\}}$. 
We can decompose ${\Theta}_{\rho\lambda}$ and $q\cdot \nabla\bar{\mu}_{\nu}$ as 
\begin{gather}
\label{Xi}
{\Theta}^{\rho\lambda}=u^{\rho}u^{\lambda}{\Pi}+2u^{\{\rho}\Pi^{\lambda\}}+\Pi^{\rho\lambda},
\\
q\cdot \nabla\bar{\mu}_{\nu}= (q\cdot u) D \bar{\mu}_{\nu}+q_{\perp}\cdot\nabla \bar{\mu}_{\nu},
\end{gather}
where
\begin{align}
\Pi &\equiv u_{\rho}u_{\lambda}{\Theta}^{\rho\lambda}=D \beta, 
\\
\Pi^{\lambda} &\equiv u^{\rho}\Delta^{\lambda\alpha}{\Theta}_{\rho\alpha}= \frac{1}{2}(\beta D u^{\lambda}+\nabla_{\perp}^{\lambda}\beta),
\\
\Pi^{\rho\lambda} &\equiv \Delta^{\rho\alpha}\Delta^{\lambda\beta}{\Theta}_{\alpha\beta}=\pi^{\rho \lambda}+\Delta^{\rho \lambda}\theta, 
\end{align}
with $\pi^{\mu\nu} \equiv \beta \nabla^{ \{ \mu}_{\perp} u^{\nu \}}-\Delta^{\mu\nu}\theta$, 
$\theta \equiv \beta \nabla_{\perp\nu}u^{\nu}/3$,
$D \equiv u \cdot \nabla$ the temporal derivative in the fluid rest frame, and
$v_{\perp}^{\mu}\equiv \Delta^{\mu\nu}v_{\nu}$ for an arbitrary vector $v^{\mu}$.

By symmetry, we expect the following constitutive relations: 
\begin{align}
\label{T_decomp}
\delta T^{(1)\mu\nu}_{(\nu)B}
&=\delta\epsilon_{B} u^{\mu}u^{\nu}-\delta p_{B_{\rm T}}\Delta^{\mu\nu}_{B}-\delta p_{B_{\rm L}}\hat{B}^{\mu}\hat{B}^{\nu}+2h_{\perp}^{\{\mu}B^{\nu\}}+2u^{\{\mu}V_{\perp}^{\nu\}},
\\
\label{J_decomp}
\delta J^{(1)\mu}_{(\nu)B}&=\delta N_{B}u^{\mu}+\sigma_{B}^{\mu\nu}B_{\nu},
\end{align}
where $\Delta^{\mu\nu}_{B}=\Delta^{\mu\nu}+\hat{B}^{\mu}\hat{B}^{\nu}$, $h_{\perp}\cdot B=0$, and $\sigma_{B}^{\mu\nu}u_{\mu}=0$.
The explicit forms of these transport coefficients read
\begin{eqnarray}
\label{deltaepsilonB}
\delta\epsilon_{B}=5\delta p_{B_{\rm T}}=-\frac{5}{3}\delta p_{B_{\rm L}}=\frac{\kappa}{6M}\int \frac{{\rm d}^3\bm q}{(2\pi)^3}\frac{f^{(\nu)}_{0,q}\big(1-f^{(\nu)}_{0,q}\big)^2}{|\bm q|\big(1-f^{(\rm e)}_{0,q}\big)}B^{\mu}\left(\beta Du_{\mu}+\nabla_{\perp\mu}\beta - \frac{\nabla_{\perp\mu}\bar{\mu}_{\nu}}{|\bm q|} \right)\,,
\end{eqnarray}
\begin{eqnarray}
\label{h}
h^{\mu}_{\perp}=-\frac{\kappa\Delta^{\mu\rho}_{B}}{30M}\int \frac{{\rm d}^3\bm q}{(2\pi)^3}\frac{f^{(\nu)}_{0,q}\big(1-f^{(\nu)}_{0,q}\big)^2}{|\bm q|\big(1-f^{(\rm e)}_{0,q}\big)}
\left(\beta Du_{\rho}+\nabla_{\perp\rho}\beta - \frac{\nabla_{\perp\rho}\bar{\mu}_{\nu}}{|{\bm q}|} \right)\,,
\end{eqnarray}
\begin{eqnarray}
\label{V}
V^{\mu}_{\perp}=\frac{\kappa}{2M}\int \frac{{\rm d}^3\bm q}{(2\pi)^3}\frac{f^{(\nu)}_{0,q}\big(1-f^{(\nu)}_{0,q}\big)^2}{|\bm q|\big(1-f^{(\rm e)}_{0,q}\big)}
\left[\frac{\Delta^{\mu\nu}}{3}\left(D\beta-\theta-\frac{D\bar{\mu}_{\nu}}{|{\bm q}|} \right)
-\frac{2}{15}\pi^{\mu\nu} \right]B_{\nu} \,,
\end{eqnarray}
\begin{eqnarray}
\label{deltaNB}
	\delta N_{B}=\frac{\kappa}{6M}\int \frac{{\rm d}^3\bm q}{(2\pi)^3}\frac{f^{(\nu)}_{0,q}\big(1-f^{(\nu)}_{0,q}\big)^2}{|\bm q|^2\big(1-f^{(\rm e)}_{0,q}\big)} B^{\mu}
	\left(\beta Du_{\mu}+\nabla_{\perp\mu}\beta - \frac{\nabla_{\perp\mu}\bar{\mu}_{\nu}}{|\bm q|} \right) \,,
\end{eqnarray}
\begin{eqnarray}
\label{sigmaB}
\sigma_{B}^{\mu\nu}=\frac{\kappa}{2M}\int \frac{{\rm d}^3\bm q}{(2\pi)^3}\frac{f^{(\nu)}_{0,q}\big(1-f^{(\nu)}_{0,q}\big)^2}{|\bm q|^2\big(1-f^{(\rm e)}_{0,q}\big)} 
\left[\frac{\Delta^{\mu\nu}}{3}\left(D\beta-\theta-\frac{D\bar{\mu}_{\nu}}{|{\bm q}|} \right)
-\frac{2}{15}\pi^{\mu\nu} \right]\,.
\end{eqnarray}
The details of the derivation are shown in Appendix~\ref{app_deltaTJhbar}. 
Here, all the temporal derivatives $D$ on the thermodynamic parameters should be replaced by spatial gradients via hydrodynamic equations shown in Sec.~\ref{sec_hydro_EOM}.

Note that $\delta N_{B}$ in Eq.~(\ref{deltaNB}) logarithmically diverges, but this may be regularized by the screening mass of the neutrino in medium. By utilizing hydrodynamic equations shown in Eq.~(\ref{hydro_EOM}), one may replace $D u^{\mu}$ by $\nabla_{\perp}^{\mu}T$ and $\nabla_{\perp}^{\mu}\bar{\mu}$ for $\bar{\mu}=(\bar{\mu}_{\rm e}, \bar{\mu}_{\rm p}, \bar{\mu}_{\rm n}, \bar{\mu}_{\nu})$
and drop the terms coupled to $D \bar{\mu}_{\nu} = O(\hbar)$ as higher order corrections in the $\hbar$ expansion.  
For simplicity, we assume $\nabla_{\perp}^{\mu}T$ and $\nabla_{\perp}^{\mu}\bar{\mu}$ are suppressed and omit $\delta\epsilon_B$, $\delta p_B$, $h^{\nu}_{\perp}$, and $\delta N_B$. The remaining terms are then given by
\begin{gather}
\label{main_V}
V^{\mu}_{\perp}=-\frac{\kappa I_{1}}{15M}\left(\pi^{\mu \nu}+\frac{5}{2}\theta \Delta^{\mu \nu}\right)B_{\nu}\,,
\\
\label{main_sigma}
\sigma_{B}^{\mu\nu}=-\frac{\kappa I_{2}}{15M}\left(\pi^{\mu\nu}+\frac{5}{2}\theta \Delta^{\mu\nu}\right)\,,
\end{gather} 
where 
\begin{align}
I_{1}&\equiv\int \frac{{\rm d}^3\bm q}{(2\pi)^3}\frac{f^{(\nu)}_{0,q}\big(1-f^{(\nu)}_{0,q}\big)^2}{|\bm q|\big(1-f^{(\rm e)}_{0,q}\big)}=\frac{T^2}{4\pi^2}\left[\frac{{\rm e}^{\bar{\mu}_{\nu}}-{\rm e}^{\bar{\mu}_{\rm e}}}{1+{\rm e}^{\bar{\mu}_{\nu}}}+(1+{\rm e}^{\bar{\mu}_{\rm e}-\bar{\mu}_{\nu}})\ln(1+{\rm e}^{\bar{\mu}_{\nu}})\right]\,, \\
I_{2}&\equiv\int \frac{{\rm d}^3\bm q}{(2\pi)^3}\frac{f^{(\nu)}_{0,q}\big(1-f^{(\nu)}_{0,q}\big)^2}{|\bm q|^2\big(1-f^{(\rm e)}_{0,q}\big)}
=\frac{T {\rm e}^{\bar{\mu}_{\nu}}(2+ {\rm e}^{\bar{\mu}_{\nu}}+ {\rm e}^{\bar{\mu}_{\rm e}})}{4\pi^2 (1+ {\rm e}^{\bar{\mu}_{\nu}})^2}\,.
\end{align}
Note that the results in Eqs.~(\ref{main_V}) and (\ref{main_sigma}) are independent of the nuclear equation of state.

Although the isoenergetic approximation may break down, it would be useful to extrapolate these results to the regime $\bar{\mu}_{\rm e}\gg 1$ and $\bar{\mu}_{\nu}\gg 1$ to obtain more compact forms, which will be used for an order of estimate in Sec.~\ref{sec_discussion}. In this regime, we find
\begin{align}
I_{1}&\approx \frac{T^2}{4\pi^2}\left[\bar{\mu}_{\nu}(1+{\rm e}^{\bar{\mu}_{\rm e}-\bar{\mu}_{\nu}})+1- {\rm e}^{\bar{\mu}_{\rm e}-\bar{\mu}_{\nu}}\right]\,,
\\
I_{2}&\approx \frac{T}{4\pi^2}(1+ {\rm e}^{\bar{\mu}_{\rm e}-\bar{\mu}_{\nu}})\,.
\end{align}
When we further assume that $\bar{\mu}_{\rm n}-\bar{\mu}_{\rm p}=\bar{\mu}_{\rm e}-\bar{\mu}_{\nu} \gg 1$ and $u^{\mu} \approx (1, {\bm v})$ with $|{\bm v}| \ll 1$, the explicit expressions for $\delta T^{(1)0 i}_{(\nu) B}$ and $\delta J^{(1)i}_{(\nu)B}$ can be simplified as
\begin{align}
\label{main_T}
\delta T^{(1)0 i}_{(\nu) B} = \delta T^{(1)i 0}_{(\nu) B} &\approx -\frac{1}{72 \pi M G_{\rm F}^2 (g_{\rm V}^2 + 3 g_{\rm A}^2)}
\frac{{\rm e}^{2\beta(\mu_{\rm n}-\mu_{\rm p})}}{n_{\rm n}-n_{{\rm p}}}
({\bm \nabla} \cdot {\bm v})\mu_{\nu} B^i\,,
\\
\label{main_J}
\delta J^{(1)i}_{(\nu)B} &\approx -\frac{1}{72 \pi M G_{\rm F}^2 (g_{\rm V}^2 + 3 g_{\rm A}^2)}
\frac{{\rm e}^{2\beta(\mu_{\rm n}-\mu_{\rm p})}}{n_{\rm n}-n_{{\rm p}}}
({\bm \nabla} \cdot {\bm v})B^i\,.
\end{align}
Note that $T^{0i} \propto B^i$ and $J^i \propto B^i$ are prohibited in usual parity-invariant matter by parity symmetry. 
However, these chiral transport become possible in the present case due to the parity-violating nature of the weak interaction.

\section{Hydrodynamic equations of motion}
\label{sec_hydro_EOM}
In this section, we present an explicit derivation of the hydrodynamic equations of motion for the system composed of nucleons, electrons, and neutrinos.
For simplicity, here we consider the hydrodynamic equations in the Lorentz covariant form, which can reduce to a nonrelativistic expression with appropriate change of variables. 
It is also sufficient to focus on the dissipationless terms for our purpose and we will ignore the dissipative terms, such as the viscosity and conductivity.

The energy-momentum tensor, vector and axial currents for electrons in local thermal equilibrium read \cite{Son:2009tf,Landsteiner:2011iq,Gao:2012ix,Yang:2018lew}
\begin{align}
\label{T}
T_{(\rm e)}^{\mu\nu}&\equiv T_{\rm R(\rm e)}^{\mu\nu}+T_{\rm L(\rm e)}^{\mu\nu}
=u^{\mu}u^{\nu}\epsilon_{(\rm e)}-p_{\rm (e)}\Delta^{\mu\nu}
+\hbar\xi_{B(\rm e)}\big(B^{\mu}u^{\nu}+B^{\nu}u^{\mu}\big)+\hbar\xi_{\omega (\rm e)}\big(\omega^{\mu}u^{\nu}+\omega^{\nu}u^{\mu}\big)\,,
\\
\label{J}
J^{\mu}_{\rm (\rm e)} &\equiv J^{\mu}_{\rm R(e)} + J^{\mu}_{\rm L(e)} =N_{\rm (\rm e)}u^{\mu}+\hbar\sigma_{B (\rm e)}B^{\mu}+\hbar\sigma_{\omega \rm (\rm e)}\omega^{\mu}\,,
\\
\label{J5}
J^{\mu}_{\rm 5(\rm e)} &\equiv J^{\mu}_{\rm R(e)} - J^{\mu}_{\rm L(e)} =N_{\rm 5(\rm e)}u^{\mu}+\hbar\sigma_{B 5(\rm e)}B^{\mu}+\hbar\sigma_{\omega \rm 5(\rm e)}\omega^{\mu}\,.
\end{align}
Here, we have
\begin{gather}
\xi_{B \rm (\rm e)}=\frac{\mu_{\rm e}\mu_{{\rm e}5}}{2\pi^2}\,,
\quad
\xi_{\omega \rm (\rm e)}=\frac{\mu_{{\rm e}5}}{3}\left(\frac{3\mu_{\rm e}^2+\mu_{{\rm e}5}^2}{\pi^2}+T^2\right)\,,
\\
N_{({\rm e})}=\frac{\mu_{\rm e}}{3}\left(\frac{3\mu_{{\rm e}5}^2+\mu_{\rm e}^2}{\pi^2}+T^2\right)\,,\quad  
N_{5({\rm e})}=\frac{\mu_{{\rm e} 5}}{3}\left(\frac{3\mu_{\rm e}^2+\mu_{{\rm e} 5}^2}{\pi^2}+T^2\right)\,,
\\
\sigma_{B\rm (e)}=\frac{\mu_{\rm e 5}}{2\pi^2}\,,\quad
\sigma_{B\rm 5(e)}=\frac{\mu_{\rm e}}{2\pi^2}\,,\quad 
\sigma_{\omega \rm (e)}=\frac{\mu_{\rm e}\mu_{\rm e5}}{\pi^2}\,,\quad
\sigma_{\omega 5(\rm e)}=\frac{\mu_{\rm e}^2+\mu_{\rm e 5}^2}{2\pi^2}+\frac{T^2}{6}\,,
\end{gather}
and, for a relativistic ideal gas of electrons,
\begin{equation}
\epsilon_{({\rm e})}=3p_{({\rm e})}=
\frac{1}{4\pi^2}({\mu}_{\rm e}^4+6{\mu}_{\rm e}^2 {\mu}_{{\rm e} 5}^2 +{\mu}_{{\rm e} 5}^4)+\frac{T^2}{2}({\mu}_{\rm e}^2+{\mu}_{{\rm e} 5}^2)+\frac{7\pi^2}{60}T^4\,.
\end{equation}
The vector and axial currents proportional to $B^{\mu}$ in Eqs.~(\ref{J}) and (\ref{J5}) are called chiral magnetic effect \cite{Vilenkin:1980fu,Nielsen:1983rb,Alekseev:1998ds,Fukushima:2008xe} and chiral separation effect \cite{Son:2004tq,Metlitski:2005pr}, respectively.

The full energy-momentum tensor in local thermal equilibrium is thus given by
\begin{eqnarray}
T^{\mu\nu}_{\rm full}=T^{\mu\nu}_{\rm mat}+T^{\mu\nu}_{(\nu)}=u^{\mu}u^{\nu}\epsilon_{\rm f}-p_{\rm f}\Delta^{\mu\nu}+\hbar\xi_{B (\rm e)}\big(B^{\mu}u^{\nu}+B^{\nu}u^{\mu}\big)
+\hbar \xi_{\omega(\rm f)} \big(\omega^{\mu}u^{\nu}+\omega^{\nu}u^{\mu}\big),
\end{eqnarray}
where $\epsilon_{\rm f}=\epsilon_{\rm (e)}+\epsilon_{(\nu)}+\epsilon_{(\rm p)}+\epsilon_{(\rm n)}$,
$p_{\rm f}=p_{\rm (e)}+p_{(\nu)}+p_{(\rm p)}+p_{(\rm n)}$, 
$\xi_{\omega(\rm f)} = \xi_{\omega({\rm e})} + \xi_{\omega(\nu)}$,
and
\begin{gather}
\epsilon_{(i)}=2 \int \frac{{\rm d}^3 {\bm q}}{(2\pi)^3}E_{q}\left(\frac{1}{{\rm e}^{\beta (E_q-\mu_i)}+1}+\frac{1}{{\rm e}^{\beta (E_q+\mu_i)}+1}\right)\,,
\\
p_{(i)}=2 \int \frac{{\rm d}^3 {\bm q}}{(2\pi)^3} \frac{|\bm q|^2}{3E_q}\left(\frac{1}{{\rm e}^{\beta (E_q-\mu_i)}+1}+\frac{1}{{\rm e}^{\beta (E_q+\mu_i)}+1}\right)\,,
\end{gather}
with $E_{q}=\sqrt{|\bm q|^2+M^2}$ for $i=\rm p, \rm n$. 
The nonequilibrium corrections, such as $\delta T^{\mu\nu}_{(\nu)}$, are responsible for higher-order gradient terms in hydrodynamic equations and are hence dropped. To be more precise, the inclusion of $\delta T^{\mu\nu}_{(\nu)}$ in hydrodynamic equations will contribute to the terms at $O(\nabla_{\perp}^2)$ for classical transport and those at $O(\nabla_{\perp}^3)$ for quantum transport, respectively, which are irrelevant in the present context. The same argument is applied to drop $\delta T_{(\rm e)}^{\mu\nu}$, $\delta J^{\mu}$, and $\delta J^{\mu}_{5}$ as nonequilibrium corrections in Eqs.~(\ref{T}), (\ref{J}), and (\ref{J5}).
The nucleon currents are given by
\begin{eqnarray}
J^{\mu}_{\rm (\rm i)}=N_{\rm (\rm i)}u^{\mu},\quad 
N_{\rm (i)}=2 \int \frac{{\rm d}^3 {\bm q}}{(2\pi)^3}\left(\frac{1}{{\rm e}^{\beta (E_q-\mu_i)}+1}-\frac{1}{{\rm e}^{\beta (E_q+\mu_i)}+1}\right),
\end{eqnarray}
for $\rm i=\rm p, \rm n$. When $\mu_{i} \gg T$, the antiparticle contributions are suppressed, and thus, $N_{(i)}=n_{i}$. 

From the lepton current conservation, anomaly relation for the axial current, electric current conservation, and baryon current conservation in Eqs.~(\ref{cons_lept})--(\ref{cons_baryon}), we have
\begin{eqnarray}
&&D (N_{\rm (e)} + N_{\rm (\nu)}) + (N_{\rm (e)} + N_{(\nu)}) \nabla\cdot u
+\hbar \nabla_{\mu}(\sigma_{B(\rm e)}B^{\mu} + \sigma_{\omega({\rm f})}\omega^{\mu})=0,
\\
&&D (N_{\rm 5(e)} - N_{\rm (\nu)}) + (N_{\rm 5(e)} - N_{(\nu)}) \nabla\cdot u
+\hbar\nabla_{\mu}(\sigma_{B5(\rm e)}B^{\mu}+\sigma_{\omega 5(\rm e)}\omega^{\mu}-\sigma_{\omega(\nu)}\omega^{\mu})=-\hbar \frac{E\cdot B}{2\pi^2}\,,
\\\label{heq_cons_echarge}
&&D (N_{\rm (p)}-N_{\rm (e)}) + (N_{\rm (p)}-N_{\rm (e)}) \nabla\cdot u - \hbar \nabla_{\mu}(\sigma_{B(\rm e)}B^{\mu} + \sigma_{\omega({\rm e})}\omega^{\mu})=0,
\\\label{heq_cons_B}
&& D (N_{\rm (p)} + N_{\rm (n)}) + (N_{\rm (p)}+ N_{\rm (n)}) \nabla\cdot u=0,
\end{eqnarray}
where $\sigma_{\omega(\rm f)} = \sigma_{\omega({\rm e})} + \sigma_{\omega(\nu)}$.
In addition, the energy-momentum conservation in Eq.~(\ref{cons_EM}) gives
\begin{eqnarray}
\nonumber
&& D \left[(\epsilon_{\rm f} + p_{\rm f}) u^{\mu} \right] +(\epsilon_{\rm f}+p_{\rm f})u^{\mu}\nabla\cdot u-u^{\mu} D p_{\rm f}-\nabla_{\perp}^{\mu}p_{\rm f}+\hbar\left[D (\xi_{\omega({\rm f})}\omega^{\mu})+ \nabla_{\nu}(\xi_{\omega({\rm f})}\omega^{\nu}u^{\mu})+\xi_{\omega({\rm f})}\omega^{\mu}\nabla\cdot u\right]
\nonumber \\
&& +\hbar\left( \omega^{\mu} \rightarrow B^{\mu}, \ \xi_{\omega({\rm f})} \rightarrow \xi_{B({\rm e})} \right)
=F^{\mu \nu}\big[(N_{\rm (p)}-N_{\rm (e)})u_{\nu}-\hbar \sigma_{B({\rm e})}B_{\nu}-\hbar \sigma_{\omega({\rm e})}\omega_{\nu}\big].
\end{eqnarray}
Here and below, ``$(\omega^{\mu} \rightarrow B^{\mu}, \ \xi_{\omega({\rm f})} \rightarrow \xi_{B({\rm e})})$" denotes the terms obtained by such replacement for the corresponding terms involving $\omega^{\mu}$ with the coefficient $\xi_{\omega({\rm f})}$.
This equation can be decomposed into the longitudinal and transverse parts with respect to $u^{\mu}$ as
\begin{eqnarray}
&& D \epsilon_{\rm f}+(\epsilon_{\rm f}+p_{\rm f})\nabla\cdot u
+\hbar\big[\xi_{\omega({\rm f})} u_{\nu} D \omega^{\nu}+ \nabla_{\mu}(\xi_{\omega({\rm f})}\omega^{\mu})\big]
+ \hbar\left( \omega^{\nu} \rightarrow B^{\nu}, \ \xi_{\omega({\rm f})} \rightarrow \xi_{B({\rm e})} \right)
\nonumber \\
&& \quad 
=\hbar \big(\sigma_{B({\rm e})}E\cdot B+\sigma_{\omega({\rm e})}E\cdot \omega\big),
\\
&&(\epsilon_{\rm f}+p_{\rm f}) D u^{\mu}-\nabla_{\perp}^{\mu}p_{\rm f}
+\hbar\left[\omega^{\mu} D \xi_{\omega({\rm f})}+ \xi_{\omega({\rm f})}\omega\cdot\nabla u^{\mu}+\xi_{\omega({\rm f})}\omega^{\mu}\nabla\cdot u
+ \xi_{\omega({\rm f})} (D \omega^{\mu} - u^{\mu} u_{\nu} D \omega^{\nu}) \right]
\nonumber \\
&& \quad +\hbar\left( \omega^{\mu} \rightarrow B^{\mu}, \ \xi_{\omega({\rm f})} \rightarrow \xi_{B({\rm e})} \right)
= E^{\mu}(N_{\rm (p)}-N_{\rm (e)})+\hbar\sigma_{\omega({\rm e})}\epsilon^{\mu \nu \alpha\beta}\omega_{\nu}B_{\alpha}u_{\beta}.
\end{eqnarray} 

More explicitly, we find
\begin{eqnarray}
&&(N_{{\rm (e)},T}+N_{(\nu),T})DT+N_{{\rm (e)},\bar{\mu}_{\rm e}}D\bar{\mu}_{\rm e}+N_{(\nu),\bar{\mu}_{\nu}}D\bar{\mu}_{\nu}+(N_{\rm (e)}+N_{(\nu)})\nabla\cdot u
\nonumber \\
&& \quad +\hbar \nabla_{\mu}(\sigma_{B(\rm e)}B^{\mu} + \sigma_{\omega({\rm f})}\omega^{\mu}) = 0,
\\\label{cons_eq_N5}
&&(N_{{\rm 5(e)},T}-N_{(\nu),T})DT+N_{\rm 5(e),\bar{\mu}_{\rm e5}}D\bar{\mu}_{\rm e5}-N_{(\nu),\bar{\mu}_{\nu}}D\bar{\mu}_{\nu} + (N_{\rm 5(e)} - N_{(\nu)}) \nabla\cdot u
\nonumber \\
&& \quad +\hbar\nabla_{\mu}(\sigma_{B5(\rm e)}B^{\mu}+\sigma_{\omega 5(\rm e)}\omega^{\mu}-\sigma_{\omega(\nu)}\omega^{\mu})=0,
\\
&&(N_{{\rm (p)},T}-N_{{\rm (e)},T})DT+N_{{\rm (p)},\bar{\mu}_{\rm p}}D\bar{\mu}_{\rm p}-N_{{\rm (e)},\bar{\mu}_{\rm e}}D\bar{\mu}_{\rm e}+(N_{\rm (p)}-N_{\rm (e)})\nabla\cdot u 
\nonumber \\
&& \quad -\hbar \nabla_{\mu}(\sigma_{B(\rm e)}B^{\mu} + \sigma_{\omega({\rm e})}\omega^{\mu}) = 0,
\\
&&(N_{{\rm (p)},T}+N_{{\rm (n)},T})DT+(N_{{\rm (p)},\bar{\mu}_{\rm p}}+N_{\rm (n),\bar{\mu}_{\rm n}})D\bar{\mu}_{\rm p}+N_{\rm (n),\bar{\mu}_{\rm n}}D(\bar{\mu}_{\rm e}-\bar{\mu}_{\rm e 5}-\bar{\mu}_{\nu})+(N_{\rm (p)}+N_{\rm (n)})\nabla\cdot u = 0,
\end{eqnarray}
and 
\begin{eqnarray}
\nonumber
&&\epsilon_{{\rm f},T}DT+(\epsilon_{{\rm f},\bar{\mu}_{\rm (e)}}+\epsilon_{{\rm f},\bar{\mu}_{\rm n}})D\bar{\mu}_{\rm (e)}
-\epsilon_{{\rm f},\bar{\mu}_{\rm n}}D\bar{\mu}_{\rm e 5}
+(\epsilon_{{\rm f},\bar{\mu}_{\rm p}}+\epsilon_{{\rm f},\bar{\mu}_{\rm n}})D\bar{\mu}_{\rm p}
+(\epsilon_{{\rm f},\bar{\mu}_{\nu}}-\epsilon_{{\rm f},\bar{\mu}_{\rm n}})D\bar{\mu}_{\nu} 
+(\epsilon_{\rm f}+p_{\rm f})\nabla\cdot u
\\
&& \quad 
+\hbar\big[ \nabla_{\mu}(\xi_{\omega({\rm f})}\omega^{\mu})-\xi_{\omega({\rm f})}\omega_{\mu}Du^{\mu} \big] 
+\hbar \left( \omega^{\mu} \rightarrow B^{\mu}, \ \xi_{\omega({\rm f})} \rightarrow \xi_{B({\rm e})} \right)
- \hbar \left(\sigma_{B({\rm e})}E\cdot B + \sigma_{\omega({\rm e})}E\cdot \omega \right) = 0,
\\\nonumber
&&(\epsilon_{\rm f}+p_{\rm f})Du^{\mu}-\nabla_{\perp}^{\mu}p_{\rm f} -E^{\mu}(N_{\rm (p)}-N_{\rm (e)})
+\hbar\big[\omega^{\mu}(\xi_{\omega({\rm f}),T}DT+\xi_{\omega({\rm f}),\bar{\mu}_{\nu}}D\bar{\mu}_{\nu}+\xi_{\omega({\rm f}),\bar{\mu}_{{\rm e}5}}D\bar{\mu}_{{\rm e}5})+ \xi_{\omega({\rm f})}(\omega\cdot\nabla u^{\mu}+\omega^{\mu}\nabla\cdot u)
\nonumber \\
&& \quad 
+ \xi_{\omega({\rm f})} (D \omega^{\mu} - u^{\mu} u_{\nu} D \omega^{\nu}) \big]
+ \hbar \left( \omega^{\mu} \rightarrow B^{\mu}, \ \xi_{\omega({\rm f})} \rightarrow \xi_{B({\rm e})} \right)
- \hbar \sigma_{\omega({\rm e})}\epsilon^{\mu\nu\alpha\beta}\omega_{\nu}B_{\alpha}u_{\beta} = 0,
\end{eqnarray}
where $F_{i,T}\equiv \partial_{T}F_{i}$ and $F_{i,\bar{\mu}_{j}}\equiv \partial_{\bar{\mu}_j}F_{i}$ correspond to the partial derivative with respect to $T$ and $\bar{\mu}_{j}$ for an arbitrary function $F_{i}(T,\bar{\mu}_j)$ and we have implemented $\mu_{\rm n}=\mu_{\rm e}+\mu_{\rm p}-\mu_{\nu}$.%
\footnote{For generality, we here used $\mu_{\rm n}=\mu_{\rm eL}+\mu_{\rm p}-\mu_{\nu}$ and took into account the contributions of ${\mu}_{\rm e 5}$. When chirality flipping occurs sufficiently rapidly, we may simply set ${\mu}_{\rm e 5}=0$ in these equations.}
Here one may further replace the combinations $D \omega^{\mu} - u^{\mu} u_{\nu} D \omega^{\nu}$ and $DB^{\mu}-u^{\mu}u_{\nu}DB^{\nu}$ by other terms via the Bianchi identities.%
\footnote{From the decomposition $\partial_{[\mu} u_{\nu]} = \epsilon_{\mu \nu \alpha \beta} u^{\alpha} \omega^{\beta} + \frac{1}{2} (u_{\mu} D u_{\nu} - u_{\nu} D u_{\mu})$ and the Bianchi identity $\epsilon^{\mu\nu\alpha\beta} \partial_{\alpha} \partial_{[\mu} u_{\nu]}=0$, we can derive \cite{Hidaka:2017auj}
\begin{equation}
D \omega^{\mu}-u^{\mu}u_{\nu}D\omega^{\nu} = - \omega^{\mu}\partial\cdot u + \omega\cdot\partial u^{\mu} - \frac{1}{2}\epsilon^{\mu\nu\alpha\beta}u_{\beta}\partial_{\nu} D u_{\alpha}\,.
\nonumber
\end{equation}
Similarly, from the decomposition ${F}_{\mu\nu}=\epsilon_{\mu \nu \alpha \beta}u^{\alpha} B^{\beta}-u_{\mu}E_{\nu}+u_{\nu}E_{\mu}$ and the Bianchi identity 
$\epsilon^{\mu\nu\alpha\beta}\partial_{\alpha}F_{\mu \nu}=0$, we have
\begin{equation}
DB^{\mu}-u^{\mu}u_{\nu}DB^{\nu} = - B^{\mu}\partial\cdot u + B\cdot\partial u^{\mu} - \epsilon^{\mu\nu\alpha\beta}\big(u_{\beta}\partial_{\nu}E_{\alpha}+u_{\nu}E_{\alpha}D u_{\beta}\big).
\nonumber
\end{equation}
}
Note that $N_{{\rm 5(e)},T}=\sigma_{B({\rm e})}=\sigma_{\omega({\rm e})}=\xi_{B({\rm e})}=\xi_{\omega({\rm e})}=0$ when $\bar{\mu}_{\rm e 5}=0$. 

Up to $O(\hbar^0)$, it is easy to show that
\begin{align}
DT&=-\frac{\epsilon_{\rm f}+p_{\rm f}}{\epsilon_{{\rm f},T}}\nabla\cdot u+O(\hbar)\,,
\\
Du^{\nu}&=\frac{\nabla_{\perp}^{\nu}p_{\rm f}}{\epsilon_{\rm f}+p_{\rm f}}+O(\hbar)
=\frac{p_{{\rm f},T}\nabla_{\perp}^{\nu}T+\sum_{\bar{\mu}} p_{{\rm f},\bar{\mu}}\nabla_{\perp}^{\nu}\bar{\mu}}{\epsilon_{\rm f}+p_{\rm f}}+O(\hbar)\,,
\end{align}
for $\bar{\mu}=(\bar{\mu}_{\rm e}, \bar{\mu}_{\rm p}, \bar{\mu}_{\rm n}, \bar{\mu}_{\nu})$,
and $D \bar \mu_i$ vanish at $O(\hbar^0)$. Here we further assumed the local charge neutrality $N_{\rm (p)}=N_{\rm (e)}$. In fact, the conservation of electric current in Eq.~(\ref{heq_cons_echarge}) is satisfied by the local charge neutrality when $\mu_{\rm e5}=0$. When $\mu_{\rm e5}\neq 0$, on the other hand, a local electric charge fluctuation can be induced at $O(\hbar)$. For relativistic ideal gases, one can find $\epsilon_{\rm f}=3p_{\rm f}$, $p_{{\rm f},T}=4p_{\rm f}/T$,  $p_{{\rm f},\bar{\mu}_{\rm e}}=N_{\rm (e)}T$, and  $p_{{\rm f},\bar{\mu}_{\nu}}=N_{(\nu)}T$, which yields 
\begin{align}
DT&=-\frac{T\nabla\cdot u}{3}+O(\hbar)\,,
\\
\label{hydro_EOM}
Du^{\mu}&=\frac{\nabla_{\perp}^{\nu}T}{T}+\frac{T}{4p_{\rm f}}\Big[(N_{\rm (e)}+\beta p_{\rm n,\bar{\mu}_{\rm n}})\nabla_{\perp}^{\nu}\bar{\mu}_{\rm e}+\beta(p_{\rm p,\bar{\mu}_{\rm p}}+p_{\rm n,\bar{\mu}_{\rm n}})\nabla_{\perp}^{\nu}\bar{\mu}_{\rm p}+(N_{\rm (\nu)}-\beta p_{\rm n,\bar{\mu}_{\rm n}})\nabla_{\perp}^{\nu}\bar{\mu}_{\nu}\Big]+O(\hbar)\,.
\end{align}
These hydrodynamic equations up to $O(\hbar^0)$ were employed to obtain the explicit expressions of $\delta T^{\mu\nu}_{(\nu)B}$ and $\delta J^{\mu}_{(\nu)B}$ in Sec.~\ref{sec_nonequilbirum_B_correcions}.

As briefly mentioned in Sec.~\ref{sec_nonequilbirum_B_correcions}, however, the magnetic field can also be involved through the temporal derivatives $D$ on the thermodynamic parameters when incorporating $\hbar$ corrections. Therefore, we need to work out the leading-order corrections in $\hbar$ expansion as well, which are shown in Appendix~\ref{app_LO_fluctuations}. For simplicity, we here set $\mu_{5{\rm e}}=0$. In this case, the magnetic field is only involved in Eq.~(\ref{cons_eq_N5}) for the hydrodynamic equations when taking $E^{\mu}=\omega^{\mu}=0$. For such a correction, one finds
\begin{eqnarray}\label{B_term_in_hydroEOM}
\hbar\nabla_{\mu}\big(\sigma_{B5(\rm e)}B^{\mu}\big)=\frac{\hbar}{2\pi^2}B^{\mu}\big(\nabla_{\perp\mu}\mu_{\rm e}-\mu_{\rm e}Du_{\mu}\big)\,.
\end{eqnarray} 
Here, we used the relation
\begin{eqnarray}
\nabla\cdot B+2E\cdot\omega+B^{\mu}Du_{\mu}=0,
\end{eqnarray}
which follows from the Bianchi identity $\nabla_{\mu}\tilde{F}^{\mu\nu}=0$ and the decomposition $\tilde{F}^{\mu\nu}=\epsilon^{\mu \nu \alpha \beta}u_{\alpha} E_{\beta}-u^{\mu}B^{\nu}+u^{\nu}B^{\mu}$. By further substituting the expression of $Du^{\mu}$ from Eq.~(\ref{hydro_EOM}), we conclude that Eq.~(\ref{B_term_in_hydroEOM}) only contains $B\cdot\nabla_{\perp}T$ and $B\cdot\nabla_{\perp}\mu$ terms, and thus, $\delta T^{\mu\nu}_{(\nu)B}$ and $\delta J^{\mu}_{(\nu)B}$ are not affected when assuming $\nabla_{\perp\mu}T=\nabla_{\perp\mu}\mu=E^{\mu}=\omega^{\mu}=0$.

\section{Discussions and outlook}
\label{sec_discussion}
Let us now consider the possible phenomenological implications of the results above. Here, we will focus especially on the neutrino momentum density $T^{i 0}_{(\nu) B}$ in Eq.~(\ref{main_T}). We can estimate the kick velocity of the core due to this contribution as
\begin{equation}
v_{\rm kick} 
\sim \frac{\delta T^{i 0}_{(\nu) B}}{\rho_{\rm core}}\,,
\end{equation}
where we assumed the homogeneous core mass density $\rho_{\rm core}$ and constant $\delta T^{i 0}_{(\nu) B}$ there for an order of estimate.
Taking $n_{\rm n} - n_{\rm p} \sim 0.1 \ {\rm fm}^{-3}$, $\mu_{\rm n} - \mu_{\rm p} \sim 100 \ {\rm MeV}$, $\mu_{\nu} \sim 100 \ {\rm MeV}$, $T \sim 10 \ {\rm MeV}$, typical length scale for the variation of the hydrodynamic variables, $L \sim 10 \ {\rm km}$, $|{\bm v}| \sim 0.01$, and $\rho_{\rm core} \sim M (n_{\rm n} + n_{\rm p})$ with $n_{\rm n} + n_{\rm p} \sim 0.1\ {\rm fm}^{-3}$, we obtain 
\begin{equation}
v_{\rm kick} \lesssim \left( \frac{B}{10^{13{\text -}14} \ {\rm G}} \right) \ {\rm km/s}\,.
\end{equation}
(The reason why this should be regarded as the upper bound will be described shortly.)
In order to account for the observed pulsar velocity $v_{\rm kick} \sim 10^2 \ {\rm km/s}$ (see, e.g., Refs.~\cite{Lyne1994,Kaspi1996,Arzoumanian2002,Hobbs2005}) solely from this contribution, the required magnetic field at the core is of order $10^{15{\text -}16} \ {\rm G}$.%
\footnote{For the previous works that attempt to explain the pulsar kick by an asymmetric neutrino emission induced by strong magnetic fields, see Refs.~\cite{Vilenkin:1995um,Horowitz:1997mk,Horowitz:1997fb,Roulet:1997sw,Lai:1997mm,Lai:1998sz,Arras:1998cv,Arras:1998mv,Goyal:1998nq,Kaminski:2014jda}.
Note that our work is the first to derive $T^{0i}_{(\nu)B}$ explicitly and systematically. The parametric dependence of $v_{\rm kick}$ here are also different from the previous results although the final order of estimate itself is comparable to Ref.~\cite{Vilenkin:1995um} among others.}
However, this estimate should be taken with care because it depends sensitively on the choice of the parameters.

From Eq.~(\ref{main_T}), one might think that for a given magnetic field, $v_{\rm kick}$ becomes arbitrarily large if $(\mu_{\rm n} - \mu_{\rm p})/T$ becomes sufficiently large. In fact, this is not the case because for a sufficiently large $(\mu_{\rm n} - \mu_{\rm p})/T$, the mean free path $\ell_{\rm mfp}$ would become larger than the typical length scale of the system, as can been seen from Eq.~(\ref{def_kappa}), where $\kappa$ increases when $(\mu_{\rm n}-\mu_{\rm p})/T$ increases. Then the assumption that neutrinos are near equilibrium would break down. This means that the kick velocity is bounded from above for a given magnetic field because of the hydrodynamic approximation.%
\footnote{Parametrically, $T^{i0}_{(\nu)B}$ may be expressed as
\begin{equation}
T^{i0}_{(\nu)B} \sim \left(\frac{\ell_{\rm mfp}}{L}\right) \frac{\mu_{\nu}^3}{M} B^i\,.
\end{equation}
Then the near-equilibrium condition of neutrinos ($L \gtrsim \ell_{\rm mfp}$) leads to the upper bound of $v_{\rm kick}$ as 
\begin{equation}
v_{\rm kick} 
\lesssim \frac{\mu_{\nu}^3 B}{M \rho_{\rm core}} \sim \left( \frac{B}{10^{13} \ {\rm G}} \right) \ {\rm km/s}\,.
\end{equation}
for $\mu_{\nu} \sim 100 \ {\rm MeV}$.}
On the other hand, the chiral radiation transport theory itself is applicable to neutrinos even far away from equilibrium, in which case such a limitation is not present. It is thus necessary to investigate the fully nonequilibrium contribution of this mechanism to provide a more realistic estimate. 

Although we have highlighted the neutrino chiral transport induced by the magnetic field near equilibrium in this paper, there are also other neutrino chiral transport induced by the vorticity and gradients of temperature and chemical potential. One expects that these chiral effects would further modify the nonlinear hydrodynamic evolution of the supernova, such as the turbulent behavior. For example, chiral/helical transport phenomena lead to the tendency toward the inverse energy cascade even in three dimensions, as analytically and numerically shown in Refs.~\cite{Yamamoto:2016xtu,Masada:2018swb} (see also Refs.~\cite{Brandenburg:2017rcb,Schober:2017cdw} in the context of the early Universe). 

We also note that neutrino chiral transport far away from equilibrium is not captured by the relaxation time approximation adopted in the present paper. In fact, even though the net momentum flux is generated for near-equilibrium neutrinos by magnetic fields, it is not guaranteed that these neutrinos can escape from the protoneutron star. This neutrino momentum flux could be canceled by the back reaction of the matter sector, and then there could be no significant emission asymmetry. The emission asymmetry might rather be caused by neutrinos outside the neutrino sphere, where the near-equilibrium approximation is not applicable. In order to see the consequences of fully nonequilibrium chiral effects, it would be eventually important to perform numerical simulations of the chiral radiation transport theory for neutrinos in the future.

\acknowledgements
This work was supported by the Keio Institute of Pure and Applied Sciences (KiPAS) project at Keio University and JSPS KAKENHI Grant No.~19K03852 and No.~20K14470 and Ministry of Science and Technology, Taiwan under Grant No. MOST-110-2112-M-001-070-MY3.

\appendix
\section{Leading-order corrections}
\label{app_LO_fluctuations}
The leading-order corrections of the energy-momentum tensor and current of neutrinos read
\begin{align}
	\delta T^{(0)\mu\nu}_{(\nu)}&=-
	\int_q \frac{4\pi\delta(q^2)}{{E_{\rm i}}}q^{\mu}q^{\nu}\tau^{(0)} f^{(\nu)}_{0,q}\big(1-f^{(\nu)}_{0,q}\big)\big(q^{\rho}q^{\lambda}{\Theta}_{\rho\lambda}-q\cdot\nabla \bar{\mu}_{\nu}\big)\,,
	\\
	\delta J^{(0)\mu}_{(\nu)}&=-
	\int_q \frac{4\pi\delta(q^2)}{{E_{\rm i}}}q^{\mu}\tau^{(0)} f^{(\nu)}_{0,q}\big(1-f^{(\nu)}_{0,q}\big)\big(q^{\rho}q^{\lambda}{\Theta}_{\rho\lambda}-q\cdot\nabla \bar{\mu}_{\nu}\big)\,.
\end{align}
By symmetry, we expect the following constitutive relations: 
\begin{align}
	\delta T^{(0)\mu\nu}_{(\nu)}
	&=\delta\epsilon^{(0)} u^{\mu}u^{\nu}-\delta p^{(0)}\Delta^{\mu\nu}+2u^{(\mu}\zeta_{\perp}^{\nu)}+\chi^{\mu\nu}_{\perp\perp},
	\\
	\delta J^{(0)\mu}_{(\nu)}&=\delta N^{(0)}u^{\mu}+j^{\mu}_{\perp},
\end{align}
where $u_{\mu}\zeta_{\perp}^{\mu}=u_{\mu}j^{\mu}_{\perp}=0$ and $u_{\mu}\chi^{\mu\nu}_{\perp\perp}=\chi^{\mu\nu}_{\perp\perp}u_{\nu}=0$. 
All these components can be computed via
\begin{gather}
\delta\epsilon^{(0)}=3\delta p^{(0)}=u_{\mu}u_{\nu}\delta T^{(0)\mu\nu}_{(\nu)}, \quad
\zeta_{\perp}^{\nu}=u_{\mu}\Delta_{\rho}^{\nu} \delta T^{(0)\mu\rho}_{(\nu)}, \quad 
\chi^{\mu\nu}_{\perp\perp} =  \Delta_{\rho}^{\mu} \Delta_{\lambda}^{\nu} \delta T^{(0)\rho \lambda}_{(\nu)}, \quad 
\\
\delta N^{(0)} = u_{\mu} \delta J^{(0)\mu}_{(\nu)}, \quad
j^{\mu}_{\perp} =  \Delta_{\rho}^{\mu} \delta J^{(0)\rho}_{(\nu)}.
\end{gather}
Their explicit expressions are given by
\begin{eqnarray}
	\delta\epsilon^{(0)}=3\delta p^{(0)}&=&-\kappa\int \frac{{\rm d}^3\bm q}{(2\pi)^3}\frac{f^{(\nu)}_{0,q}\big(1-f^{(\nu)}_{0,q}\big)^2}{1-f^{(\rm e)}_{0,q}}
\left(\hat{q}^{\rho}\hat{q}^{\lambda}{\Theta}_{\rho\lambda}-\frac{\hat{q}\cdot\nabla \bar{\mu}_{\nu}}{|\bm q|}\right)
\nonumber \\
	&=&-\kappa\int \frac{{\rm d}^3\bm q}{(2\pi)^3}\frac{f^{(\nu)}_{0,q}\big(1-f^{(\nu)}_{0,q}\big)^2}{1-f^{(\rm e)}_{0,q}}\left(D \beta - \theta -\frac{D\bar{\mu}_{\nu}}{|\bm q|}\right)\,,
\end{eqnarray} 
\begin{eqnarray}\nonumber
\zeta_{\perp}^{\nu}&=&-\kappa\int \frac{{\rm d}^3\bm q}{(2\pi)^3}\frac{f^{(\nu)}_{0,q}\big(1-f^{(\nu)}_{0,q}\big)^2}{1-f^{(\rm e)}_{0,q}}
\hat{q}^{\nu}_{\perp}\left(\hat{q}^{\rho}\hat{q}^{\lambda}{\Theta}_{\rho\lambda}-\frac{\hat{q}\cdot\nabla \bar{\mu}_{\nu}}{|\bm q|}\right)
\\
&=&\kappa\int \frac{{\rm d}^3\bm q}{(2\pi)^3}\frac{f^{(\nu)}_{0,q}\big(1-f^{(\nu)}_{0,q}\big)^2}{3\big(1-f^{(\rm e)}_{0,q}\big)}
\left(\beta Du^{\nu}+\nabla^{\nu}_{\perp}\beta-\frac{\nabla_{\perp}^{\nu}\bar{\mu}_{\nu}}{|\bm q|}\right)\,,
\end{eqnarray}
\begin{eqnarray}\nonumber
\chi^{\mu\nu}_{\perp\perp}&=&-\kappa\int \frac{{\rm d}^3\bm q}{(2\pi)^3}\frac{f^{(\nu)}_{0,q}\big(1-f^{(\nu)}_{0,q}\big)^2}{1-f^{(\rm e)}_{0,q}}
\hat{q}_{\perp}^{\mu}\hat{q}^{\nu}_{\perp}\left(\hat{q}^{\rho}\hat{q}^{\lambda}{\Theta}_{\rho\lambda}-\frac{\hat{q}\cdot\nabla \bar{\mu}_{\nu}}{|\bm q|}\right)
\\
&=&\kappa\int \frac{{\rm d}^3\bm q}{(2\pi)^3}\frac{f^{(\nu)}_{0,q}\big(1-f^{(\nu)}_{0,q}\big)^2}{1-f^{(\rm e)}_{0,q}}
\left[\frac{\Delta^{\mu\nu}}{3}\left(D\beta-\theta-\frac{D\bar{\mu}_{\nu}}{|\bm q|}\right) - \frac{2}{15}\pi^{\mu\nu} \right]\,,
\end{eqnarray}
\begin{eqnarray}\nonumber
\delta N^{(0)}&=&-\kappa\int \frac{{\rm d}^3\bm q}{(2\pi)^3}\frac{f^{(\nu)}_{0,q}\big(1-f^{(\nu)}_{0,q}\big)^2}{|\bm q|\big(1-f^{(\rm e)}_{0,q}\big)}
\left(\hat{q}^{\rho}\hat{q}^{\lambda}{\Theta}_{\rho\lambda}-\frac{\hat{q}\cdot\nabla \bar{\mu}_{\nu}}{|\bm q|}\right)
\\
&=&-\kappa\int \frac{{\rm d}^3\bm q}{(2\pi)^3}\frac{f^{(\nu)}_{0,q}\big(1-f^{(\nu)}_{0,q}\big)^2}{|\bm q|\big(1-f^{(\rm e)}_{0,q}\big)}
\left(D\beta-\theta-\frac{D\bar{\mu}_{\nu}}{|\bm q|}\right)\,,
\end{eqnarray}
\begin{eqnarray}\nonumber
j^{\mu}_{\perp}&=&-\kappa\int \frac{{\rm d}^3\bm q}{(2\pi)^3}\frac{f^{(\nu)}_{0,q}\big(1-f^{(\nu)}_{0,q}\big)^2}{|\bm q|\big(1-f^{(\rm e)}_{0,q}\big)}
\hat{q}_{\perp}^{\mu} \left(\hat{q}^{\rho}\hat{q}^{\lambda}{\Theta}_{\rho\lambda}-\frac{\hat{q}\cdot\nabla \bar{\mu}_{\nu}}{|\bm q|}\right)
\\
&=&\kappa\int \frac{{\rm d}^3\bm q}{(2\pi)^3}\frac{f^{(\nu)}_{0,q}\big(1-f^{(\nu)}_{0,q}\big)^2}{3|{\bm q}|\big(1-f^{(\rm e)}_{0,q}\big)}
\hat{q}_{\perp}^{\mu} \left(\beta Du^{\mu}+\nabla_{\perp}^{\mu}\beta-\frac{\nabla_{\perp}^{\mu}\bar{\mu}_{\nu}}{|\bm q|} \right)\,.
\end{eqnarray}

\section{Evaluations of $\delta T^{(1)\mu\nu}_{(\nu)B}$ and $\delta J^{(1)\mu}_{(\nu)B}$}
\label{app_deltaTJhbar}
In this appendix, we provide the derivations of Eqs.~(\ref{deltaepsilonB})--(\ref{sigmaB}).
Given the decomposition of $\Theta^{\mu\nu}$ in Eq.~(\ref{Xi}), one may write $\delta T^{(1)\mu\nu}_{(\nu)B}$ and $\delta J^{(1)\mu}_{(\nu)B}$ as
\begin{align}
\delta T^{(1)\mu\nu}_{B}&=-
\int\frac{{\rm d}^3\bm q}{(2\pi)^3}\hat{q}^{\mu}\hat{q}^{\nu}\tau^{(0)} \frac{\hat{q}\cdot B}{2M}f^{(\nu)}_{0,q}\big(1-f^{(\nu)}_{0,q}\big)
|\bm q| \left[ \left(\Pi- \frac{D\bar{\mu}_{\nu}}{|\bm q|} \right)+\hat{q}_{\perp}^{\rho}\left(2\Pi_{\rho}-\frac{\nabla_{\perp\rho}\bar{\mu}_{\nu}}{|\bm q|} \right)
+\Pi_{\rho\lambda}\hat{q}^{\rho}_{\perp}\hat{q}^{\lambda}_{\perp}\right]\,,
\\
\delta J^{(1)\mu}_{(\nu)B}&=-
\int\frac{{\rm d}^3\bm q}{(2\pi)^3}\hat{q}^{\mu}\tau^{(0)}\frac{\hat{q}\cdot B}{2M}f^{(\nu)}_{0,q}\big(1-f^{(\nu)}_{0,q}\big)
\left[ \left(\Pi- \frac{D\bar{\mu}_{\nu}}{|\bm q|} \right)+\hat{q}_{\perp}^{\rho}\left(2\Pi_{\rho}-\frac{\nabla_{\perp\rho}\bar{\mu}_{\nu}}{|\bm q|} \right)
+\Pi_{\rho\lambda}\hat{q}^{\rho}_{\perp}\hat{q}^{\lambda}_{\perp}\right]\,.
\end{align}
All the relevant coefficients of the decompositions in Eqs.~(\ref{T_decomp}) and (\ref{J_decomp}) can be calculated via
\begin{gather}
\delta\epsilon_{B}=u_{\mu}u_{\nu}\delta T^{(1)\mu\nu}_{(\nu)B},\quad 
\delta p_{B_{\rm T}}=-\frac{1}{2}\Delta_{B\mu\nu}\delta T^{(1)\mu\nu}_{(\nu)B},
\quad
\delta p_{B_{\rm L}}=-\hat{B}_{\mu}\hat{B}_{\nu}\delta T^{(1)\mu\nu}_{(\nu)B},
\nonumber\\
h_{\perp}^{\mu}=-\frac{1}{|{\bm B}|}\hat{B}_{\rho} (\Delta_B)_{\nu}^{\mu} \delta T^{(1)\rho\nu}_{(\nu)B},\quad
V_{\perp}^{\mu}=\Delta^{\mu}_{\rho} u_{\nu} \delta T^{(1)\rho\nu}_{(\nu)B},
\end{gather}
and
\begin{gather}
\delta N_{B}=u_{\mu} \delta J^{(1)\mu}_{(\nu)B},
\quad
\sigma_{B}^{\mu\nu}B_{\nu} = \Delta^{\mu}_{\rho} \delta J^{(1)\rho}_{(\nu)B},
\end{gather}
where we used $\Delta_{B\mu\nu}\Delta_{B}^{\mu\nu}=2$.

When evaluating the integrals, we also use the following useful relations for an arbitrary function ${\cal F}(|{\bm q}|)$:
\begin{gather}
\int\frac{{\rm d}^3\bm q}{(2\pi)^3} \hat{q}_{\perp}^{\mu}\hat{q}_{\perp}^{\nu} {\cal F}(|{\bm q}|)
= \int\frac{{\rm d}^3\bm q}{(2\pi)^3} \left[{z}^2 \hat{B}^{\mu}\hat{B}^{\nu}-\frac{(1-{z}^2)}{2}\Delta^{\mu\nu}_{B} \right] {\cal F}(|{\bm q}|)\,,
\\
\int\frac{{\rm d}^3\bm q}{(2\pi)^3} \hat{q}_{\perp}^{\mu}\hat{q}_{\perp}^{\nu}\hat{q}^{\rho}_{\perp} {\cal F}(|{\bm q}|)
= \int\frac{{\rm d}^3\bm q}{(2\pi)^3} 
\left[{z}^3\hat{B}^{\mu}\hat{B}^{\nu}\hat{B}^{\rho}-\frac{{z}(1-{z}^2)}{2}
\big(\hat{B}^{\mu}\Delta_{B}^{\nu\rho}+\hat{B}^{\nu}\Delta_{B}^{\mu\rho}+\hat{B}^{\rho}\Delta_{B}^{\mu \nu}\big)\right] {\cal F}(|{\bm q}|) \,,
\end{gather}
where ${z}\equiv - \hat B \cdot \hat q_{\perp} = \hat{\bm B}\cdot \hat{\bm q}$. 
One then finds
\begin{eqnarray}\nonumber
\delta\epsilon_{B}&=&-
\int\frac{{\rm d}^3\bm q}{(2\pi)^3}\tau^{(0)} \frac{\hat{q}\cdot B}{2M}f^{(\nu)}_{0,q}\big(1-f^{(\nu)}_{0,q}\big)
|\bm q| \left[ \left(\Pi- \frac{D\bar{\mu}_{\nu}}{|\bm q|} \right)+\hat{q}_{\perp}^{\rho}\left(2\Pi_{\rho}-\frac{\nabla_{\perp\rho}\bar{\mu}_{\nu}}{|\bm q|} \right)
+\Pi_{\rho\lambda}\hat{q}^{\rho}_{\perp}\hat{q}^{\lambda}_{\perp}\right]
\\\nonumber
&=&
\int\frac{{\rm d}^3\bm q}{(2\pi)^3}\tau^{(0)} \frac{{z}^2}{2M}f^{(\nu)}_{0,q}\big(1-f^{(\nu)}_{0,q}\big)
B^{\rho}(2|\bm q|\Pi_{\rho}-\nabla_{\perp\rho}\bar{\mu}_{\nu})
\\
&=&-\frac{\kappa}{6M}\int \frac{{\rm d}^3\bm q}{(2\pi)^3}\frac{f^{(\nu)}_{0,q}\big(1-f^{(\nu)}_{0,q}\big)^2}{|\bm q|\big(1-f^{(\rm e)}_{0,q}\big)}
B^{\mu} \left(\beta Du_{\mu}+\nabla_{\perp\mu}\beta - \frac{\nabla_{\perp\mu}\bar{\mu}_{\nu}}{|{\bm q}|}\right)\,,
\end{eqnarray}
\begin{eqnarray}\nonumber
\delta p_{B_{\rm T}}&=&-\frac{1}{2}\int\frac{{\rm d}^3\bm q}{(2\pi)^3}(1-{z}^2)\tau^{(0)} \frac{\hat{q}\cdot B}{2M}f^{(\nu)}_{0,q}\big(1-f^{(\nu)}_{0,q}\big)
|\bm q| \left[ \left(\Pi- \frac{D\bar{\mu}_{\nu}}{|\bm q|} \right)+\hat{q}_{\perp}^{\rho}\left(2\Pi_{\rho}-\frac{\nabla_{\perp\rho}\bar{\mu}_{\nu}}{|\bm q|} \right)
+\Pi_{\rho\lambda}\hat{q}^{\rho}_{\perp}\hat{q}^{\lambda}_{\perp}\right]
\\\nonumber
&=&\frac{1}{2}\int\frac{{\rm d}^3\bm q}{(2\pi)^3}\tau^{(0)} \frac{(1-{z}^2){z}^2}{2M}f^{(\nu)}_{0,q}\big(1-f^{(\nu)}_{0,q}\big)B^{\rho}(2|\bm q|\Pi_{\rho}-\nabla_{\perp\rho}\bar{\mu}_{\nu})
\\
&=&\frac{1}{5}\delta\epsilon_{B}\,,
\end{eqnarray}
\begin{eqnarray}
\nonumber
\delta p_{B_{\rm L}}&=&\int\frac{{\rm d}^3\bm q}{(2\pi)^3}{z}^2\tau^{(0)} \frac{\hat{q}\cdot B}{2M}f^{(\nu)}_{0,q}\big(1-f^{(\nu)}_{0,q}\big)
|\bm q| \left[ \left(\Pi- \frac{D\bar{\mu}_{\nu}}{|\bm q|} \right)+\hat{q}_{\perp}^{\rho}\left(2\Pi_{\rho}-\frac{\nabla_{\perp\rho}\bar{\mu}_{\nu}}{|\bm q|} \right)
+\Pi_{\rho\lambda}\hat{q}^{\rho}_{\perp}\hat{q}^{\lambda}_{\perp}\right]
\\\nonumber
&=&-\int\frac{{\rm d}^3\bm q}{(2\pi)^3}\tau^{(0)} \frac{{z}^4}{2M}f^{(\nu)}_{0,q}\big(1-f^{(\nu)}_{0,q}\big)B^{\rho}(2|\bm q|\Pi_{\rho}-\nabla_{\perp\rho}\bar{\mu}_{\nu})
\\
&=&-\frac{3}{5}\delta\epsilon_{B}\,,
\end{eqnarray}
\begin{eqnarray}\nonumber
h^{\mu}_{\perp}&=&
\int\frac{{\rm d}^3\bm q}{(2\pi)^3}{z}^2 \tau^{(0)} \frac{\hat{q}^{\mu}_{\perp}-{z}\hat{B}^{\mu}}{2M}f^{(\nu)}_{0,q}\big(1-f^{(\nu)}_{0,q}\big)
|\bm q| \left[ \left(\Pi- \frac{D\bar{\mu}_{\nu}}{|\bm q|} \right)+\hat{q}_{\perp}^{\rho}\left(2\Pi_{\rho}-\frac{\nabla_{\perp\rho}\bar{\mu}_{\nu}}{|\bm q|} \right)
+\Pi_{\rho\lambda}\hat{q}^{\rho}_{\perp}\hat{q}^{\lambda}_{\perp}\right]
\\\nonumber
&=&-\frac{1}{2}\int\frac{{\rm d}^3\bm q}{(2\pi)^3} \tau^{(0)} \frac{{z}^2(1-{z}^2)}{2M}f^{(\nu)}_{0,q}\big(1-f^{(\nu)}_{0,q}\big)\Delta^{\mu\rho}_{B}(2|\bm q|\Pi_{\rho}-\nabla_{\perp\rho}\bar{\mu}_{\nu})
\\
&=&-\frac{\kappa}{2M}\int \frac{{\rm d}^3\bm q}{(2\pi)^3}\frac{f^{(\nu)}_{0,q}\big(1-f^{(\nu)}_{0,q}\big)^2}{|\bm q|\big(1-f^{(\rm e)}_{0,q}\big)}\frac{\Delta^{\mu\rho}_{B}}{15}
\left(\beta Du_{\rho}+\nabla_{\perp\rho}\beta - \frac{\nabla_{\perp\rho}\bar{\mu}_{\nu}}{|\bm q|}\right)\,,
\end{eqnarray}
\begin{eqnarray}\nonumber
V_{\perp}^{\mu}&=&-
\int\frac{{\rm d}^3\bm q}{(2\pi)^3}\hat{q}^{\mu}_{\perp}\tau^{(0)} \frac{\hat{q}\cdot B}{2M}f^{(\nu)}_{0,q}\big(1-f^{(\nu)}_{0,q}\big)
|\bm q| \left[ \left(\Pi- \frac{D\bar{\mu}_{\nu}}{|\bm q|} \right)+\hat{q}_{\perp}^{\rho}\left(2\Pi_{\rho}-\frac{\nabla_{\perp\rho}\bar{\mu}_{\nu}}{|\bm q|} \right)
+\Pi_{\rho\lambda}\hat{q}^{\rho}_{\perp}\hat{q}^{\lambda}_{\perp}\right]
\\\nonumber
&=&
\int\frac{{\rm d}^3\bm q}{(2\pi)^3}\tau^{(0)} \frac{{z}^2}{2M}f^{(\nu)}_{0,q}\big(1-f^{(\nu)}_{0,q}\big)
|\bm q| \bigg[B^{\mu}\left(\Pi- \frac{D\bar{\mu}_{\nu}}{|\bm q|}\right)
-\left((1-{z}^2)\Pi^{\mu\rho}B_{\rho}+\frac{(3-5{z}^2)}{2}B^{\mu}\hat{B}_{\nu}\Pi^{\nu\rho}\hat{B}_{\rho}\right)
\\\nonumber
&&-\frac{(1-{z}^2)}{2}\Pi^{\rho}_{\rho}B^{\mu}\bigg]
\\\nonumber
&=&\frac{\kappa}{2M}\int \frac{{\rm d}^3\bm q}{(2\pi)^3}\frac{f^{(\nu)}_{0,q}\big(1-f^{(\nu)}_{0,q}\big)^2}{|\bm q|\big(1-f^{(\rm e)}_{0,q}\big)}
\left[\frac{\Delta^{\mu\nu}}{3}\left(D\beta-\theta-\frac{D\bar{\mu}_{\nu}}{|{\bm q}|}\right)
-\frac{2}{15}\pi^{\mu\nu}\right]B_{\nu}\,,
\end{eqnarray}
\begin{eqnarray}\nonumber
\delta N_{B}&=&-
\int\frac{{\rm d}^3\bm q}{(2\pi)^3}\tau^{(0)} \frac{\hat{q}\cdot B}{2M}f^{(\nu)}_{0,q}\big(1-f^{(\nu)}_{0,q}\big)
\left[ \left(\Pi- \frac{D\bar{\mu}_{\nu}}{|\bm q|} \right)+\hat{q}_{\perp}^{\rho}\left(2\Pi_{\rho}-\frac{\nabla_{\perp\rho}\bar{\mu}_{\nu}}{|\bm q|} \right)
+\Pi_{\rho\lambda}\hat{q}^{\rho}_{\perp}\hat{q}^{\lambda}_{\perp}\right]
\\\nonumber
&=&
\int\frac{{\rm d}^3\bm q}{(2\pi)^3}\tau^{(0)} \frac{{z}^2}{2M}f^{(\nu)}_{0,q}\big(1-f^{(\nu)}_{0,q}\big)
B^{\mu}\left(2\Pi_{\mu}-\frac{\nabla_{\mu} \bar{\mu}_{\nu}}{|\bm q|} \right)
\\
&=&\frac{\kappa}{6M}\int \frac{{\rm d}^3\bm q}{(2\pi)^3}\frac{f^{(\nu)}_{0,q}\big(1-f^{(\nu)}_{0,q}\big)^2}{|\bm q|^2\big(1-f^{(\rm e)}_{0,q}\big)}
B^{\mu} \left(\beta Du_{\mu}+\nabla_{\perp\mu}\beta - \frac{\nabla_{\perp\mu}\bar{\mu}_{\nu}}{|{\bm q}|}\right)\,,
\end{eqnarray}
\begin{eqnarray}\nonumber
\sigma_{B}^{\mu\nu}B_{\nu}&=&-
\int\frac{{\rm d}^3\bm q}{(2\pi)^3}\hat{q}^{\mu}_{\perp} \tau^{(0)} \frac{\hat{q} \cdot B}{2M}f^{(\nu)}_{0,q}\big(1-f^{(\nu)}_{0,q}\big)
\left[ \left(\Pi- \frac{D\bar{\mu}_{\nu}}{|\bm q|} \right)+\hat{q}_{\perp}^{\rho}\left(2\Pi_{\rho}-\frac{\nabla_{\perp\rho}\bar{\mu}_{\nu}}{|\bm q|} \right)
+\Pi_{\rho\lambda}\hat{q}^{\rho}_{\perp}\hat{q}^{\lambda}_{\perp}\right]
\\\nonumber
&=&
\int\frac{{\rm d}^3\bm q}{(2\pi)^3}\tau^{(0)} \frac{{z}^2}{2M}f^{(\nu)}_{0,q}\big(1-f^{(\nu)}_{0,q}\big)
\left[B^{\mu} \left(\Pi-\frac{D\bar{\mu}_{\nu}}{|{ \bm q}|} \right)
-(1-{z}^2)\pi^{\mu\rho}B_{\rho}-\frac{5(1-{z}^2)}{2}\theta B^{\mu}\right]
\\
&=&\frac{\kappa}{2M}\int \frac{{\rm d}^3\bm q}{(2\pi)^3}\frac{f^{(\nu)}_{0,q}\big(1-f^{(\nu)}_{0,q}\big)^2}{|\bm q|^2\big(1-f^{(\rm e)}_{0,q}\big)}
\left[\frac{\Delta^{\mu\nu}}{3}\left(D\beta-\theta-\frac{D\bar{\mu}_{\nu}}{|{\bm q}|}\right)
-\frac{2}{15}\pi^{\mu\nu}\right]B_{\nu}\,.
\end{eqnarray}
\bibliography{neutrino_near_equilbrium_v2.bbl}
\end{document}